\documentclass[a4paper,11pt]{article}
\pdfoutput=1 

\usepackage{jheppub,amsmath} 

\usepackage[T1]{fontenc} 

\usepackage{times}
\usepackage{graphicx}
\usepackage{amssymb}
\usepackage{xspace}
\usepackage{caption}
\usepackage{subcaption}
\usepackage{slashed}
\usepackage{verbatim}
\usepackage[utf8]{inputenc}
\usepackage{breqn}

\def\be{\begin{eqnarray}}
\def\ee{\end{eqnarray}}

\newcommand{\mt}[1]{\textrm{\tiny #1}}

\def\pt{{p_\mt{T}}}

\def\aSC{{\kappa_{\rm sc}}}
\def\lres{L_{\rm res}}
\def\rres{R_{\rm res}}

\def\x{{\bf x}}

\newcommand{\pythia}{{\sc Pythia}\xspace}

\def\mpt{\slashed{p}^\parallel_\mt{T}}

\title{Resolution Effects in the Hybrid Strong/Weak Coupling Model}

\author[a]{Zachary Hulcher,}
\author[b]{Daniel Pablos,}
\author[a]{Krishna Rajagopal}

\affiliation[a]{Center for Theoretical Physics, Massachusetts Institute of Technology, Cambridge, MA 02139 USA}

\affiliation[b]{
Department of Physics, McGill University, 3600 University Street, Montr\'eal, QC, H3A 2T8, Canada}

\emailAdd{zhulcher@mit.edu}

\emailAdd{pablosd@physics.mcgill.ca}

\emailAdd{krishna@mit.edu}

\preprint{{\footnotesize MIT-CTP-4920}}

\abstract{

Within the context of a hybrid strong/weak coupling model of jet quenching, we study the consequences of the fact that the plasma produced in a heavy ion collision cannot resolve the substructure of a collimated parton shower propagating through it with arbitrarily fine spatial resolution.
We introduce a screening length parameter, $\lres$, proportional to the inverse of the local temperature in the plasma, 
estimating a range for the value of the proportionality constant via comparing 
weakly coupled QCD calculations and holographic calculations appropriate in strongly coupled plasma.
We then modify the hybrid model so that when a parton in a jet shower splits, its two offspring are initially treated as unresolved, and are only treated as two separate partons losing energy independently after they are separated by a distance $\lres$.  This modification delays the quenching of partons with intermediate energy, resulting in the survival of more hadrons in the final state with $p_T$ in the several GeV range.
We analyze the consequences of different choices for the value of the resolution length, $\lres$,
and
demonstrate that introducing a nonzero $\lres$
results in modifications to the jet shapes and jet fragmentations functions, as it makes it more probable for particles carrying a small fraction of the jet energy at larger angles from the jet axis to survive their passage through the quark-gluon plasma.
These effects are, however, small in magnitude, something that we confirm via checking for effects on missing-$p_T$ observables.
}

\begin{document} 
\maketitle
\flushbottom

\section{Introduction}

High energy heavy ion collisions at the Relativistic Heavy Ion Collider (RHIC) and the Large Hadron Collider (LHC) provide a key window into the dynamics and properties of droplets of the hot matter that filled the microseconds old universe, called quark-gluon plasma (QGP).
Experiments at these facilities have demonstrated that in the experimentally accessible range of temperatures, up to several times hotter than the crossover temperature at which cooling QGP becomes ordinary hadronic matter, 
droplets of QGP 
exhibit strong collective 
phenomena~\cite{Adcox:2004mh,Arsene:2004fa,Back:2004je,Adams:2005dq, Aamodt:2010pa,ATLAS:2011ah,Chatrchyan:2012ta},
with the dynamics of the rapid expansion and cooling of the initially lumpy droplets produced in 
the collisions
successfully described by the equations of  relativistic viscous hydrodynamics~\cite{Huovinen:2001cy,Teaney:2001av,Hirano:2005xf,Romatschke:2007mq,Luzum:2008cw,Schenke:2010rr,hiranoLHC,Gale:2012rq,Shen:2014vra,
Shen:2014nfa,
Bernhard:2016tnd}.
The ratio of the shear viscosity, $\eta$, to the entropy density, $s$, serves as a benchmark, because in a weakly coupled plasma, $\eta/s\propto  1/g^4$ (with $g$ the gauge coupling), meaning that this ratio is large, 
whereas
$\eta/s = 1/4\pi$ in the high temperature phase (conventionally called the plasma phase even though in reality it is a liquid) of any gauge theory that has a dual gravitational description in the limit of strong coupling and large number of colors~\cite{Policastro:2001yc,Buchel:2003tz,Kovtun:2004de}. Comparisons between hydrodynamic calculations of, and experimental measurements of, anisotropic flow in heavy ion collisions indicate that the QGP in QCD has an $\eta/s$ that is comparable to, and in particular not much larger than $1/4\pi$, meaning 
that QGP itself is a strongly coupled liquid.

The discovery that QGP is a strongly coupled liquid challenges us to find experimental means to probe its structure and properties.  The only probes that we have available are those
produced in the same heavy ion collisions in which the droplets of QGP themselves are produced.
Here we shall focus entirely on the use of high transverse momentum jets, produced at the moment of the collision in initial hard scatterings, as probes.
As a partonic jet shower propagates through the strongly coupled
plasma created in a heavy ion collision, the partons lose energy and momentum 
as a consequence of their strong interactions with the plasma, creating a wake in the plasma.  
These interactions lead to a reduction in the  jet energy (or quenching) 
and to modifications  of the properties of jets produced in heavy ion
collisions relative to those of their counterparts produced in proton-proton collisions, that propagate in vacuum.
By pursuing a large suite of jet measurements, the different LHC collaborations have observed 
strong modification of different jet observables  in heavy ion collisions~\cite{Aad:2010bu,Chatrchyan:2011sx,Chatrchyan:2012nia,Chatrchyan:2012gt,Chatrchyan:2012gw,Aad:2012vca,Raajet:HIN,Aad:2013sla,Chatrchyan:2013kwa,Abelev:2013kqa,Chatrchyan:2013exa,Chatrchyan:2014ava,Aad:2014wha,Aad:2014bxa,Adam:2015ewa,Adam:2015doa,Aad:2015bsa,HIN-16-002,Khachatryan:2015lha,Khachatryan:2016erx,Khachatryan:2016tfj,Khachatryan:2016jfl,Aaboud:2017bzv,Acharya:2017goa,Sirunyan:2017jic,Aaboud:2017eww},  
making jets promising probes for medium diagnostics. 
The first experimental constraints on jet quenching came from hadronic measurements at RHIC~\cite{Adcox:2001jp,Adler:2002xw,Adler:2002tq}. Analyses of
jets themselves and their modification 
are also being performed at RHIC~\cite{Ploskon:2009zd,Perepelitsa:2013faa,Adamczyk:2013jei,Jacobs:2015srw,Adamczyk:2016fqm,Adamczyk:2017yhe} and are one of the principal scientific goals of the
planned sPHENIX detector~\cite{Adare:2015kwa}.

A complete theoretical description of the processes by which jets are modified via passage through QGP remains challenging, because it is a multi-scale problem.
On the one hand, the production of jets and the processes via which an initial hard parton fragments into a shower
are weakly coupled hard processes.
On the other hand, the interaction of jets with the medium, the dynamics of softer components
within jets, and the dynamics of the wake produced in the medium by the jets are all sensitive to strongly coupled dynamics of the plasma at scales of order its temperature. 
One class of approaches is based upon assuming that resummed weakly coupled analyses can be
applied almost throughout.
(See Refs.~\cite{Jacobs:2004qv,CasalderreySolana:2007zz,Majumder:2010qh,Ghiglieri:2015zma,Blaizot:2015lma,Qin:2015srf} for reviews and 
Refs.~\cite{Zapp:2008af,Zapp:2008gi,Armesto:2009fj,Schenke:2009gb,Lokhtin:2011qq,Zapp:2012ak,Zapp:2013vla,Zapp:2013zya} 
for
Monte Carlo tools for analyzing jet observables that are being developed based upon these approaches.)
However, the observation that QGP  is a strongly coupled liquid tells us that physics at scales of order its temperature
is governed by strong coupling dynamics.
This realization has lead to many fruitful connections between the physics of the QCD plasma 
and the gauge/gravity duality~\cite{Maldacena:1997re}. This technique allows us rigorous and quantitative 
access to nonperturbative, strongly coupled, physics 
in a large family of non-abelian gauge theory plasmas
that have a dual holographic description in terms of a black hole spacetime in a gravitational theory with one higher dimension. 
Although the current formulation of the duality has not been shown to apply to QCD, the study of the plasmas in gauge theories that do have a holographic description has led to many insights into the dynamics of hot deconfined matter in QCD. (See 
Refs.~\cite{CasalderreySolana:2011us,DeWolfe:2013cua,Chesler:2015lsa} for reviews.) 
Within this context, there have been many interesting studies that address varied aspects of the interaction between high energy probes and strongly 
coupled plasma~\cite{Herzog:2006gh,Liu:2006ug,CasalderreySolana:2006rq,Gubser:2006bz,Liu:2006nn,Liu:2006he,Gubser:2006nz,Chernicoff:2006hi,CasalderreySolana:2007qw,Chesler:2007an,Gubser:2007ga,Chesler:2007sv,Hofman:2008ar,Gubser:2008as,Hatta:2008tx,Dominguez:2008vd,Chesler:2008wd,Chesler:2008uy,D'Eramo:2010ak,Arnold:2010ir,Arnold:2011qi,Arnold:2011vv,Chernicoff:2011xv,Chesler:2011nc,Arnold:2012uc,Arnold:2012qg,Chesler:2013urd,Ficnar:2013wba,Ficnar:2013qxa,Chesler:2014jva,Rougemont:2015wca,Chesler:2015nqz,Casalderrey-Solana:2015tas,Rajagopal:2016uip,Brewer:2017dwd}.
None of these approaches, however, can treat the intrinsically weakly coupled processes of jet production and fragmentation, since in all examples that are currently accessible via gauge/gravity duality the gauge theory is strongly coupled in the ultraviolet, rather than asymptotically free.

To address the multi-scale dynamics of QCD jets in strongly coupled plasma more fully, in 
Refs.~\cite{Casalderrey-Solana:2014bpa,Casalderrey-Solana:2015vaa,Casalderrey-Solana:2016jvj} 
two of us together with coauthors have introduced and developed a phenomenological hybrid 
strong/weak coupling approach to analyzing jet quenching. In this approach, different physical processes of relevance for the interaction of developing jet showers with the quark gluon plasma are treated differently:
the production and evolution of the jet shower is treated perturbatively, 
because the physics governing these processes is expected to be weakly coupled, while 
the interaction between each of the partons formed in the shower with the medium 
is assumed to follow the rate of energy loss of an energetic quark in strongly coupled
plasma obtained via holographic calculations in Refs.~\cite{Chesler:2014jva,Chesler:2015nqz}.

In this paper we remedy a lacuna in the hybrid model of Refs.~\cite{Casalderrey-Solana:2014bpa,Casalderrey-Solana:2015vaa,Casalderrey-Solana:2016jvj}.  In the model as 
developed to now, when a parton in the jet shower splits, the two resulting offspring partons 
are treated as if they separately and independently lose energy to the plasma from the moment
of the splitting.  This cannot be correct: when the two offspring are separated by a distance that
is $\ll 1/T$, there is no way that a medium with temperature $T$ can resolve them and respond
to them separately.
Our goal in this paper is to assess how important it is to include the finite resolving power of
the medium in the hybrid strong/weak coupling model by introducing a simple extension
of the model allowing us to treat
the two offspring as a single unresolved parton, losing energy as the parent parton
was,  until the offspring are separated
by a distance $\lres$  that we shall specify.  This is certainly an oversimplification, as we shall
explain, but it allows us to characterize the consequences of finite resolution via introducing only a single 
new parameter into the model and calculating its effects on several key observables.
We find that introducing a nonzero resolution length, $\lres$, brings the predictions of the 
hybrid model for the jet fragmentation function and the jet shape into slightly better agreement
with data, but the effects are small in magnitude.
We also check whether these effects are sufficient to bring the model predictions for 
the so-called missing-$p_T$ observables into agreement with data in the $p_T$-range of several GeV where discrepancies have been found~\cite{Casalderrey-Solana:2016jvj}, and in this case
find no significant improvement.
This lends indirect support to the alternative explanation suggested in Ref.~\cite{Casalderrey-Solana:2016jvj} for these particular discrepancies, namely that the wake created in the liquid medium by a jet 
does not fully thermalize.

We begin in Section~\ref{HybSum} by providing a brief review of the hybrid
model of Refs.~\cite{Casalderrey-Solana:2014bpa,Casalderrey-Solana:2015vaa,Casalderrey-Solana:2016jvj}. 
In Section~\ref{Resin}, we then describe the modification to the model
that we shall investigate, introducing a resolution length, $\lres$, that must be of order the Debye length in the plasma, as we shall discuss. 
Partons separated by less than $\lres$ cannot be resolved by the medium and
so should lose energy in the hybrid model as one parton (with their combined energy and momentum) does.  
When two colored partons propagating through the plasma are sufficiently close to each other,  this should be physically indistinguishable from
a single parton with the same total energy, momentum, and color charge propagating through the
same plasma.
Analogous observations have been made in at least two other contexts: in weakly coupled analyses of how QCD dipoles with varying sizes radiate gluons as they propagate through the plasma, where the analogous observation arises as a consequence of quantum interference~\cite{MehtarTani:2010ma,MehtarTani:2011tz,CasalderreySolana:2011rz}, 
{and in an
analysis of strongly coupled proxies for pairs of jets in holography~\cite{Casalderrey-Solana:2015tas}.
Once partons are separated by more than $\lres$, however, they behave in the plasma (and in particular should lose energy) as separate partons.  
In the hybrid model as developed in Refs.~\cite{Casalderrey-Solana:2014bpa,Casalderrey-Solana:2015vaa,Casalderrey-Solana:2016jvj}, $\lres=0$: when a parton splits in two, its offspring start losing energy as if they were independent immediately.
In Section~\ref{Implement}, we describe
our simplified implementation of the effects of resolution, $\lres\neq 0$, in the hybrid model.
In Section~\ref{Effect}, we 
assess the consequences of introducing a nonzero $\lres$ for several observables 
calculated previously in the hybrid model, namely
hadronic jet shapes, hadronic fragmentation functions, and two different missing-$p_T$ observables.
Introducing resolution effects means that partons in the shower that have a relatively low energy ``hide'' for a while, since as far as the plasma is concerned they remain lumped together with their higher energy parent parton for a longer time. This makes it less likely that they lose all their
energy to the medium and more likely that they survive as components of the jet until the jet
hadronizes. This
results in  small modifications to both the fragmentation function (enhancing it
for hadrons carrying a modest fraction of the jet energy) and the jet shape (enhancing
it at moderate angles, where these hadrons are found).
We close in Section~\ref{Discuss} with a discussion and a look ahead.

\section{\label{HybSum} Brief Summary of the Hybrid Strong/Weak Coupling Model}

The hybrid strong/weak coupling model is a phenomenological approach to the multi-scale problem one encounters in describing 
jet quenching phenomena  in heavy-ion collisions. The production of high energy jets is under good theoretical control through perturbative QCD calculations due to the high virtuality scale which characterizes the process; furthermore, these processes occur at  very early time scales, much earlier 
than the formation of the QGP. 
This high virtuality relaxes by successive splittings, 
which are again well described in perturbative QCD, via DGLAP (Dokshitzer-Gribov-Lipatov-Altarelli-Parisi) evolution.
Indeed, the medium temperature $T$, which is 
the relevant energy scale characterizing the strongly coupled QGP liquid, 
is much smaller than the virtuality carried by the energetic partons 
in the fragmenting shower, 
and for this reason we will assume that 
(even 
while the  partons in the shower lose energy to, and 
exchange momentum with, the plasma
as they propagate through it) 
the structure of the 
branching shower is unmodified.
This simplifying assumption is one of the bases for the hybrid model. It is a good approximation during the early stages of the shower when the virtuality of the partons
in the shower is much larger than any virtuality of order $T$ which is injected by the medium. In the later stages of the shower, it is a simplifying assumption that could be revisited in future work.

As they pass through the plasma, the partons in the jet shower will transfer energy and momentum to the medium at a rate that should be 
described without assuming weak coupling, since 
the typical momentum exchange of such interactions is of the order of the temperature $T$. 
In the hybrid model, these processes are modeled by assuming that the rate of energy loss
takes on the same form as that for a light quark passing through the strongly 
coupled plasma of ${\cal N}=4$ supersymmetric Yang-Mills (SYM) theory with a large number of colors $N_c$,
the simplest strongly coupled gauge theory with a dual, or holographic, gravitational description.
The light quark is described via a string, and energy loss corresponds to parts of the string being
absorbed by a black-hole horizon in one higher dimension~\cite{Chesler:2008uy,Chesler:2014jva,Chesler:2015nqz}.
This geometric description of parton energy loss yields new intuition and new qualitative insights into the otherwise challenging strongly coupled dynamics of the jet/plasma interaction. 
For example~\cite{Chesler:2015nqz,Rajagopal:2016uip}, in this description if two jets have the same energy the one with the wider opening
angle loses more energy, similar to a phenomenon that also arises at weak coupling~\cite{CasalderreySolana:2012ef,Milhano:2015mng,Escobedo:2016jbm} and in  the hybrid model~\cite{Casalderrey-Solana:2016jvj}, in those contexts because 
wider jets contain more partons than narrower ones, and that means that jet quenching results in a population of jets in which narrower jets predominate~\cite{Milhano:2015mng,Rajagopal:2016uip}.   The holographic calculations of Refs.~\cite{Chesler:2014jva,Chesler:2015nqz} also yield a specific analytic form for the rate of parton energy loss, 
that has then been applied parton-by-parton to the partons in a jet shower in the hybrid model:
\be
\label{CR_rate}
\left. \frac{d E_{\rm parton}}{dx}\right|_{\rm strongly~coupled}= - \frac{4}{\pi} E_{\rm in} \frac{x^2}{x_{\rm therm}^2} \frac{1}{\sqrt{x_{\rm therm}^2-x^2}}\,, 
\ee 
where $E_{\rm in}$ is the initial energy of the parton before it loses any energy to the plasma and where $x_{\rm therm}$ is the jet thermalization distance, or stopping distance,  and is given by
\be
\label{CR_xstop}
\quad \quad x_{\rm therm}= \frac{1}{2\aSC}\frac{E_{\rm in}^{1/3}}{T^{4/3}}\ ,
\ee
with $\aSC$ a parameter that depends on the 't Hooft coupling, $g^2 N_c$, as well as on  
details of the gauge theory and of how the energetic parton is prepared.
In the hybrid model~\cite{Casalderrey-Solana:2014bpa,Casalderrey-Solana:2015vaa,Casalderrey-Solana:2016jvj},
$\aSC$ is treated as a free parameter that has been fixed by comparing the model predictions for one measured quantity 
to data, as described below.
The parameter $\aSC$ is the single free parameter in the hybrid model; it controls the
magnitude of $dE/dx$, the rate of parton energy loss
whose form is given by (\ref{CR_rate}), and
as such its value affects the hybrid model predictions for every observable.
The fitted value of $\aSC$ turns out to correspond to a value of the thermalization 
length $x_{\rm therm}$ for the liquid QGP of QCD produced in heavy ion collisions that is
about 3 to 4 times longer than that for the strongly coupled ${\cal N}=4$ SYM plasma with the same temperature,
a result that is not unreasonable given the larger number of degrees of freedom in 
the latter theory~\cite{Casalderrey-Solana:2014bpa,Casalderrey-Solana:2015vaa,Casalderrey-Solana:2016jvj}.

In this Section, we provide a brief description of 
the hybrid model and how it has been used to calculate how various jet observables
are modified in heavy ion collisions relative to proton-proton collisions, as a consequence of the passage of the partons in the jet shower through
the plasma produced in a heavy ion collision.
A more detailed account of the base model may be found in Refs.~\cite{Casalderrey-Solana:2014bpa,Casalderrey-Solana:2015vaa}. 
Equation~(\ref{CR_rate}) gives the rate of energy loss of each parton 
as it traverses the plasma and, as discussed in Ref.~\cite{Casalderrey-Solana:2014bpa}, 
the lifetime of each parton from when it is created at a splitting to when it itself splits is taken to be 
\be
\label{timetolive}
\quad \quad \tau= 2\frac{E}{Q^2},
\ee
with $E$  the energy of the parton and $Q$ its virtuality. 
The factor of 2 connects this equation in the soft limit with the 
standard expression for the formation time. 
This prescription, which assigns a space-time structure to the parton shower in terms of the formation times, 
is supported by certain weak coupling computations, for example the medium induced two gluon inclusive emission calculated in Ref.~\cite{Casalderrey-Solana:2015bww} in which it is seen that 
the second emission is delayed by precisely the formation time of the parton produced at the
first emission.
Equations~(\ref{CR_rate}) and (\ref{timetolive}) together provide the basis for the hybrid model. 

In the hybrid model, we begin by taking parton showers generated by simulating 
collisions in \pythia~\cite{Sjostrand:2007gs}
and giving the partons in these showers  lifetimes according to Equation (\ref{timetolive}). This
gives the shower a structure in space and time. 
We then place the point of origin of the event generated by \pythia at a location in the transverse
plane of a heavy ion collision, choosing the location proportional to the number of collisions at that
point in the transverse plane.
For simplicity, we let the shower evolve without modification for an initial proper time that we 
take to be $\tau=0.6$~fm, and then turn on a medium as described by a hydrodynamic 
calculation. (See 
Refs.~\cite{Casalderrey-Solana:2014bpa,Casalderrey-Solana:2015vaa} for the full specification of the hydrodynamic backgrounds that we have used, and references. Here we shall
follow 
Refs.~\cite{Casalderrey-Solana:2015vaa,Casalderrey-Solana:2016jvj}.) 
It would certainly be worth investigating the effects of additional energy loss before
the formation of a hydrodynamic medium, but we shall not attempt this here.
As the parton shower develops from $\tau=0.6$~fm onward, we track the 
position in space and time of each parton and apply the energy loss $dE/dx$
from  Eq.~(\ref{CR_rate}), using the local $T$ taken from the hydrodynamic background at that point
in space and time in the expression (\ref{timetolive}) for $\x_{\rm therm}$.  We then turn energy loss off for each parton in the shower
when the temperature at its location drops below a temperature that we vary between 145 
and 170 MeV~\cite{Casalderrey-Solana:2014bpa}. 
If we wish to compute hadronic observables (like fragmentation functions, jet shapes, and mssing-$p_T$ observables) as opposed to calorimetric observables like jet $R_{AA}$, we hadronize the events with \pythia's Lund String Model.  We then run the 
FastJet anti-$k_t$ algorithm~\cite{Cacciari:2008gp,Cacciari:2011ma} to reconstruct jets, choosing the same value for the reconstruction parameter $R$ 
used in whichever experimental analysis we wish to compare to.
In essence, choosing a value of $R$ in the anti-$k_t$ 
reconstruction algorithm says that clusters of energy should be 
reconstructed as part of a single jet if they are within $R$ of each other in 
the $\eta-\phi$ plane, with $\eta$ and $\phi$ being pseudorapidity and azimuthal angle, respectively.

In order to compare this model to data, we first have to fit its one parameter, $\aSC$, 
which we do by comparing the model predictions 
for jet $R_{AA}$ to experimental data.  Specifically, we look at $R_{AA}$ for jets reconstructed with anti-$k_t$ reconstruction parameter 
$R=0.3$ that have transverse momentum in the range 100~GeV$<p_T< 110$~GeV 
and pseudorapidities in the range $-2<\eta<2$ 
in the $10\%$ most central Pb-Pb collisions with collision energy $\sqrt{s_{NN}}=2.76$ TeV. 
We fix the range for the parameter $\aSC$ such that the hybrid
model results for this specified jet $R_{AA}$ measurement matches the CMS data point~\cite{Raajet:HIN,Khachatryan:2016jfl},
including its error bar. 

As discussed in 
Refs.~\cite{Casalderrey-Solana:2014bpa, Casalderrey-Solana:2016jvj, Casalderrey-Solana:2015vaa}, 
the hybrid model has enjoyed considerable
success in modeling the dijet asymmetry, dijet imbalance, photon-Jet observables, and Z-Jet observables, most recently in the remarkably successful comparison made by CMS of hybrid model  predictions for various
photon-jet observables to a suite of new CMS measurements~\cite{HIN-16-002} 
that appeared after the predictions.
All these observables are sensitive to the 
energy loss, but not to modifications of the shape of the jets.
However, even after adding two key effects that are missing from this base framework (transverse broadening and the backreaction of the medium, see below) in Ref.~\cite{Casalderrey-Solana:2016jvj}, 
the hybrid model predictions for jet shapes, fragmentation functions, and two missing-$p_T$ observables do not agree quantitatively with data.  Several explanations for this were posited in Ref.~\cite{Casalderrey-Solana:2016jvj}, including the possibility that leaving out the effects of resolution was over-quenching moderate energy partons in the shower as well as the possibility that
the wake in the plasma was not fully thermalized as assumed in Ref.~\cite{Casalderrey-Solana:2016jvj}.  Here, we shall provide an assessment of the first possibility.

Before turning to the effects of resolution, though, we close this Section with a brief review
of the two effects added in Ref.~\cite{Casalderrey-Solana:2016jvj}.
Transverse momentum broadening is the effect that as the partons in the jet
shower propagate through the plasma they receive kicks tranverse to their direction of motion.
As appropriate in a strongly coupled plasma, or for propagation over a sufficiently long distance
through a weakly coupled plasma, the transverse momentum picked up by each parton after travelling a distance $L$ is 
Gaussian distributed with a width given by $\sqrt{\hat q L}$, where $\hat q = KT^3$. $K$ is a constant at strong coupling and a constant $\propto g^4$ up to a logarithm at weak coupling.  In principle, $K$ is a new parameter that, like $\aSC$, should be fit to data.  However, it turns out~\cite{Casalderrey-Solana:2016jvj} that jet observables measured to date are relatively insensitive to $K$ because the softer partons that would receive a  noticeable kick are in fact much more affected by energy loss, so much so that the angular-narrowing effects of energy loss (softer partons at larger angles getting fully quenched; wider jets losing more energy than narrower ones) 
substantially dominate over the angular-broadening coming from transverse kicks. 
Because its visible effects are small, in the present paper  we shall set $K=0$ throughout.

The backreaction of the medium, which is to say the wake left behind in the plasma by the
passing jet, is another physical effect that cannot be ignored. The partons in the jet lose energy {\it
and momentum}, which must both be deposited into the medium in the form of a wake.  By momentum conservation, this wake must have a net momentum in the jet direction.
This means that after hadronization when a reconstruction algorithm is used to reconstruct jets, the reconstructed jets must include some hadrons that come from the hadronization of the wake in
the plasma, as well as hadrons coming from the jet itself.  This is unavoidable, and must
be taken into account in the hybrid model, or in any other model for jet quenching, if one wishes
to compare to experimental data on the soft particles in jets and/or the structure of jets at larger angles.  In experimental data, there is no way even in principle to separate which of the hadrons that
are reconstructed as a jet in fact come from the wake in the plasma.
In Ref.~\cite{Casalderrey-Solana:2016jvj},
the transfer of energy and momentum from hard jet modes to soft plasma modes
is assumed to result in a perturbation to the hydrodynamic background that can be treated to linear order, which in turn becomes a linearized perturbation to hadron spectra after hadronization.} This perturbation is assumed to have thermalized fully, subject to momentum conservation. After hadronization according to the Cooper-Frye prescription, these assumptions then permit 
an analytic computation of the contribution to the spectrum of particles in the final state coming
from the boosted and heated wake in the plasma that is given by~\cite{Casalderrey-Solana:2016jvj}
\begin{dmath}
	\label{backreaction}
	E\frac{d\Delta N}{d^3p}=\frac{m_T}{32\pi T^5}\cosh(y-y_j)\exp\left[\frac{-m_T}{T}\cosh(y-y_j)\right]\times\left\{p_{\bot}\Delta P_{\bot}\cos(\phi-\phi_j)+\frac{m_T \Delta M_T}{3}\cosh(y-y_j)\right\},
\end{dmath}
where $p_T$, $m_T$, $\phi$, and $y$ are the transverse momentum, transverse mass, azimuthal angle, and rapidity of the emitted thermal particles coming from the wake in the plasma due to the passage of a jet whose azimuthal angle and rapidity are $\phi_j$ and $y_j$.

With a brief look back at the hybrid model in hand, we now turn to our discussion of the resolution length and its incorporation into the hybrid model framework.

\section{\label{Resin} The Resolution Length within QGP}

When one parton in a jet shower splits into two, the two offspring will initially be arbitrarily close together.
Initially, therefore, they should continue to interact with the plasma as if they were still 
the single parent parton.
Only after they have separated by some distance that we refer to as the resolution length
and denote by  $\lres$
will they interact with the plasma independently, as two separate color-charged partons each losing energy and momentum to the
plasma.
Depending on the opening angle of the splitting, it may take considerably longer than $\lres/c$ for
them to separate from each other by a distance $\lres$.  
Note that, in reality, the physical processes via which the plasma goes from ``seeing'' one parent to resolving
two offspring may be complex and/or quantum mechanical.  Our goal here is not a complete description of the resolution process.
Rather, in the next Section we shall implement a simplified prescription in which one effective parent parton suddenly becomes two offspring once the offspring are separated by $\lres$.  We shall
describe the implementation, including its attendant simplifications, there. In this section,
we start by asking what the reasonable range of values for $\lres$ should be.

Because the plasma screens color charges that are separated from each other by more than the Debye, or screening, length $\lambda_D\equiv 1/m_D$, with $m_D$ the Debye mass, we know that $\lres$ must be of order $\lambda_D$ or shorter.  When two partons are separated by a distance that is greater than $\lambda_D$, they must engage with the plasma independently.
It is also reasonable to expect that when  two partons are much closer than $\lambda_D$ to
each other, well within their own Debye spheres, the medium will not be able to resolve them,
but in the case of a weakly coupled plasma this needs further thought, see below.

For the strongly coupled plasma of any theory with a holographic dual, 
$\lambda_D$ is of order $1/T$
and hence is of order the shortest relevant length scale that characterizes the plasma.
We therefore conclude that in a strongly coupled plasma, $\lres$ is indeed of order  $\lambda_D$, not
shorter.  For the
specific case of strongly coupled ${\cal N}=4$ SYM theory, the Debye length
has been computed holographically in Ref.~\cite{Bak:2007fk} and is given by 
$\lambda_D\approx 0.3/(\pi T)$.  Because the QCD plasma has fewer degrees of freedom,
it is reasonable to expect that in temperature regimes where it is strongly coupled, its $\lambda_D$
is larger by a factor of a few, making it reasonable to estimate that strongly coupled QCD
plasma has $\lres \sim 1/(\pi T)$.

What about for a weakly coupled plasma?  Here $m_D \sim g T$ meaning that in the strict weak coupling limit the Debye length $\lambda_D$ is parametrically larger than $1/T$, and we should consider whether $\lres$ could be parametrically smaller than $\lambda_D$.
One way to gain intuition is to consider the perturbative computations of medium-induced gluon radiation from a dipole, the so-called antenna problem discussed for example in Refs.~\cite{MehtarTani:2010ma,MehtarTani:2011tz,CasalderreySolana:2011rz,Casalderrey-Solana:2015bww}.  In these calculations, the gluon radiation is
controlled entirely by the summed charge of the dipole if the opening angle of the 
antenna $\theta_{\rm ant}$ is small enough such that $\theta_{\rm ant}L \ll \lres$, with $L$ the thickness  of the medium through which the dipole propagates.  That is, if the charges in the dipole never separate by more than $\lres$, as a consequence of quantum interference 
they radiate as if they were a single charge, unresolved by 
the medium. In the case of a thin medium, this $\lres$ is indeed given by $\lambda_D$~\cite{Casalderrey-Solana:2015bww}.
If $\theta_{\rm ant}$ is larger and the legs of the antenna separate at a faster rate, the interference rapidly fades away, and the gluon emission spectrum is dominantly that for independent emission 
from each of the individual antenna charges separately.  So, for a thin medium at weak coupling, $\lres \sim \lambda_D$.  

For a thick medium at weak coupling, however, the argument is more subtle because quantum interference over the course of multiple soft interactions with the medium as a hard parton 
propagates for a distance $L$ introduces a new length scale, $1/\sqrt{\hat q L}$. (See, for example, Refs.~\cite{CasalderreySolana:2011rz,CasalderreySolana:2012ef}. For a review of earlier work
going back to Ref.~\cite{Baier:1996sk} in which this scale appears, see Ref.~\cite{CasalderreySolana:2007zz}.) 
Taking $\hat q \sim g^4 T^3$, we see that $1/\sqrt{\hat q L}$ is shorter than 
$\lambda_D\sim1/(gT)$ only if the hard parton has propagated over an $L$ that is 
longer than of order $1/(g^2 T)$. 
(Note that $L>{\cal O}(1/(g^2 T))$ is also the criterion for the medium to be considered thick.)
If we return to the antenna
problem but now for a thick medium,
if $\theta_{\rm ant} > {\cal O}(g)$ then the two legs of the antenna will reach a separation of order $\lambda_D$ before they have traveled a distance of order $1/(g^2 T)$, meaning that 
they will be resolved when their
separation is of order $\lambda_D$
and the
scale $1/\sqrt{\hat q L}$ never becomes relevant.
In the strict weak coupling limit, 
this argument suffices.
We shall take this as sufficient evidence to proceed using the assumption that $\lres\sim \lambda_D$,
as this is valid in a strongly coupled plasma as well as in a weakly coupled plasma 
except for the case where
$\theta_{\rm ant} < {\cal O}(g)$ in a plasma that is both thick and weakly coupled, a
case that can be 
returned to in future work.

In a weakly coupled plasma, the textbook result for $m_D$ can be found, for example,
in Ref.~\cite{DEramo:2012uzl} and is given by $m_D^2=\frac{3}{2}g^2 T^2$ for QCD with $N_c=N_f=3$.  Taking $\alpha_{QCD}\equiv g^2/(4\pi)$ to lie within the broad range from $\frac{1}{8}$ to  $\frac{1}{2}$ corresponds to taking $\lambda_D$ to lie within the broad range from $2.0/(\pi T)$
to $1.0/(\pi T)$.  

Reflecting these considerations, it seems that 
\begin{equation}
\frac{1}{\pi T} \lesssim \lres \lesssim \frac{2}{\pi T}
\end{equation}
is a reasonable range of expectation for $\lres$ in either a strongly coupled QCD plasma or a weakly coupled QCD plasma.
In implementing the effects of resolution in the hybrid model, we shall define
\begin{equation}
\lres=\frac{\rres}{\pi T}\, ,
\end{equation}
treating $\rres$ as a new parameter in the model.  
With this parametrization, we ensure that $\lres \propto 1/T$ in the hybrid model, with $T$ the local temperature at the position of each parton in the shower at a given time.
We will explore the consequences of choosing the proportionality constant
$\rres=1$ and $\rres=2$, spanning the reasonable range  for the resolution length in either strongly coupled or weakly coupled QCD plasma.  We shall also explore the consequences
of choosing $\rres=5$, corresponding to an unphysically large value of the resolution length. 
We do so in order to check the robustness of our results, i.e.~for the purpose of confirming that none of the results we report are hypersensitive to the value of $\lres$ that we choose.

\section{\label{Implement}Implementation of the Effects of Resolution within the Hybrid Model}

In this Section we will describe our implementation of the effects of resolution 
within the hybrid strong/weak coupling model, whose main features have already been described in Section~\ref{HybSum}. 
Noting that the hybrid model is itself only a model, our goal is not a complete description of the physical processes via which two offspring partons produced in a splitting go from being unresolved to being fully resolved and losing energy separately.  Our goal is a simplified prescription that suffices to gauge the magnitude of the effects of including resolution on several physical observables
that are of interest and where it is reasonable to expect some effects, with a magnitude to be determined.

Until now, the hybrid model has been built upon a space-time picture for the development of 
a parton shower that is based solely on the formation time argument (\ref{timetolive}),
such that a specific parton splits into its two offspring once the formation time for that parton
has passed. This process iterates until all partons are on-shell.  Each parton in the shower
loses energy independently, according to (\ref{CR_rate}), from the moment that it is produced
in a splitting.  Including the effects of resolution serves to modify the space-time structure
of the shower that the medium sees, responds to, and affects.  
Because it takes time before the offspring from each splitting in the shower are
separated by $\lres$ and resolved,
the 
moments in time when the number of shower partons seen by the medium
 increases are delayed.
The medium now sees the shower as a collection of 
\emph{effective partons}, each of which is either a single parton that has separated from
its siblings by a distance $\lres$ and is losing energy independently, or a collection of offspring that are still being treated together as a single
not-yet-resolved entity, and as such are all losing the same fractional energy.
There are certainly issues with this oversimplified prescription.  The description we have given
sounds simple and straightforward only at first hearing; implementing it for a branching shower in a dynamically evolving
medium necessitates resolving several important ambiguities.
The most obvious one
is that when we state that a single effective parton at some location becomes two resolved partons 
at two separated locations at the moment
when the two offspring partons are separated by $\lres$, we must specify in which
Lorentz frame we are making this statement.  This choice has effects and is arbitrary, and the fact that we must
make it reflects the oversimplified nature of our treatment of the process of resolution.  As we shall see below, our implementation requires us to pick the same Lorentz frame in our treatment of
all splittings, and for this reason we
shall choose to apply our prescription in the laboratory frame even though in the hybrid model
we compute the energy loss of each parton, or effective parton, in its own local fluid rest frame.  
We describe further ambiguities, and the attendant simplifying 
assumptions that we make, below.

\begin{figure}[t]
	\centering
	\includegraphics[width=.9\textwidth]{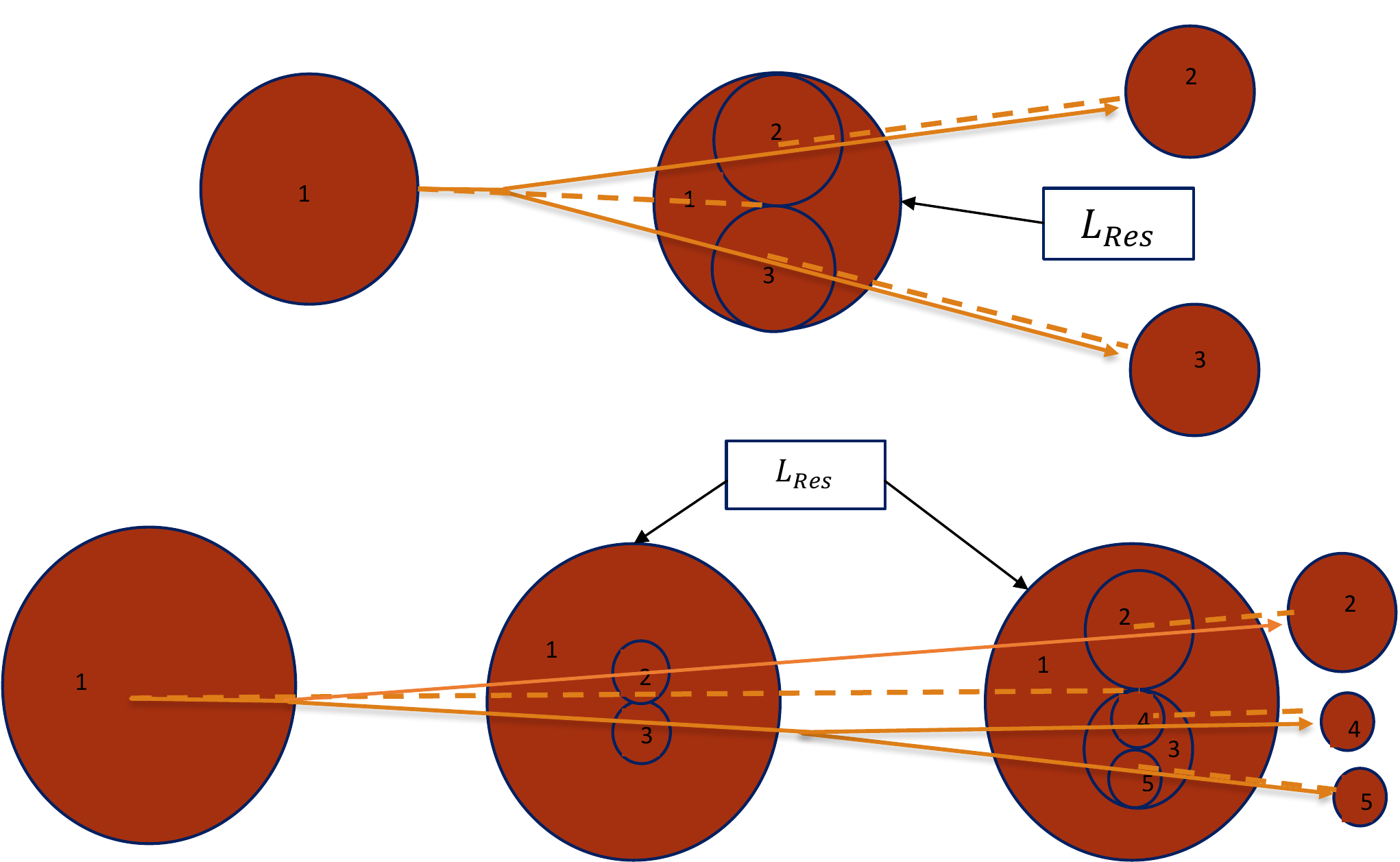}
	\caption{\label{Fig:tree} A simplified picture illustrating how we implement the effects
	of resolution. Solid lines indicate the particles in the parton shower, without any consideration of resolution. Dashed lines indicate the new shower seen by the medium after taking into account that it can only resolve offspring partons as distinct when they have separated by a distance $\lres$.
	In the top diagram, particle 1 propagates as a single effective parton after it splits until its offspring, particles 2 and 3, separate by $\lres$. In the bottom diagram, 
particles 2 and 3 have not separated by $\lres$ before particle 3 splits into particles 4 and 5.	
It is the later resolution of particles 4 and 5 that ``breaks'' the effective parton: particle 1 resolves into particles 2, 4, and 5 when this happens. 
	}
\end{figure}

As the first and simplest example, consider the picture in the upper panel of Fig.~\ref{Fig:tree}. 
Here we find depicted the splitting of parent particle $m_1$ into sibling particles $s_2$ and $s_3$. 
The solid orange lines show the ``true'' trajectory of the partons in the system, determined by the formation time argument only. If instead we require $s_2$ and $s_3$ to be spatially separated by more than $\lres$ in order to be treated as individual resolved objects from the point of view of the medium, we get the ``effective'' trajectories shown by the dashed orange lines: 
the medium perceives that $m_1$ has lived for a time $\tau_r$, which is longer than the formation time $\tau_f$, and sees particles $s_2$ and $s_3$ simultaneously pop up at $\tau_r$ at separated spatial locations. 
Therefore, when we apply our energy loss prescription (\ref{CR_rate}) to this system, 
we will need to quench $m_1$, with its summed color charge, momentum vector and (decreasing) total energy, 
as if it had propagated until $\tau_r$ before splitting.  And, after its energy loss is propagated
to its two now separated offspring at $\tau_r$, each of its offpsring in turn loses energy according
to (\ref{CR_rate}) starting from $\tau_r$.  
Only after $\tau_r$ do we have two effective partons, 
 losing energy independently.

In order to determine $\tau_r$, we need to check at each time step after the splitting
at $\tau_f$ whether the spatial separation between the offspring $s_2$ and $s_3$ (a non-local quantity) is greater than $\lres$, a quantity that depends on the local temperature of the plasma. 
As we already noted, we need to decide in which Lorentz frame to make this check:
we evaluate the spatial distance $\Delta x$ between $s_2$ and $s_3$ at the same time, but at the same
time in which frame?
Once we consider a shower with multiple splittings, the simplest way of avoiding
reintroducing ambiguities from the nonocality of our prescription is to
make the same choice of frame for all splittings. We choose to use 
the laboratory frame.  
Furthermore, we need to decide 
which temperature to use in defining $\lres$: we choose the temperature at the location
of the effective parton, $m_1$, (the dashed orange line) at the same laboratory frame time  
at which we are measuring $\Delta x$, the spatial separation between $s_2$ and $s_3$.
$\tau_r$ is then the laboratory frame time at which $\Delta x$ has become equal to $\lres$.
This completes the specification of our simplified implementation of resolution --- for the simple case depicted in the top panel of Fig.~\ref{Fig:tree}.

We turn now to a more involved, and in fact more realistic situation, where further splittings occur before the first two offspring are resolved from each other. 
This situation is illustrated in the lower panel of Fig.~\ref{Fig:tree}. 
Consider $m_1$ splitting into $s_2$ and $s_3$, where $s_3$ splits into $d_4$ and $d_5$ before $\Delta x_{23}$,  the spatial separation between $s_2$ and $s_3$,  becomes 
greater than $\lres$. In this example, the first separation to exceed its relevant $\lres$ is the one between $d_4$ and $d_5$, namely $\Delta x_{45}$, and not $\Delta x_{23}$. 
Of course, the reader will observe that before $\Delta x_{45}>\lres$, $s_2$ and $d_5$ may 
separate by $\Delta x_{25}>\lres$, and one could imagine this causing the resolution. 
The reason why we don't attempt to specify a prescription based upon this is
that when $d_5$ resolves from $s_2$, $d_4$ is still unresolved both from $s_2$ and $d_5$, meaning that there would be an ambiguity regarding the arrangement of these three partons into effective partons: should $d_4$ belong to an effective parton together with $s_2$, or to one together 
with $d_5$? 
In order to avoid these and several other ambiguities arising from our phenomenological 
approach to the problem, we prescribe that the resolution check is to be done only between closest relatives, building effective partons from sibling particles that can in turn 
be combined with {\it their} siblings to form even larger effective partons.
In the example shown in the lower panel of Fig.~\ref{Fig:tree},
we never consider the distances $\Delta x_{24}$ or $\Delta x_{25}$.
As long as $\Delta x_{45}<\lres$, we treat $d_4$ and $d_5$ as an unresolved effective
parton. And, as long as the distance between this effective parton and $s_2$ is less than $\lres$,
the three particle system behaves as a single effective parton, following the trajectory of 
$m_1$. 
This persists
until the moment when $\Delta x_{45}>\lres$, where particles $s_2$, $d_4$ and $d_5$ suddenly appear. Thus, in this example particle $s_3$ is never seen by the plasma as 
an independent parton losing energy in its own right.
Indeed, resolving $d_4$ and $d_5$ has caused $s_2$ and $s_3$ to be resolved
since, at that point, $s_3$ no longer exists as an effective parton that could be paired with $s_2$.
The algorithm that we have described via this example is not the only possible way in
which to implement the effects of resolution.  Its principle virtue is that it is fully specified, with
no remaining ambiguities, and as such will meet our needs. Recall that our goal
is a simplified prescription that we can use to gauge the magnitude of the effect of
including a nonzero $\lres$ on various observables.

Let us now walk through the logic of the algorithm  that we shall follow to modify the 
space-time structure of any shower, generalizing beyond the examples depicted in Fig.~\ref{Fig:tree}.
First, we fully reconstruct the parton shower by calculating each particle's initial and final position, initial momentum, parent particle, creation time, and splitting (or finishing) time, exactly as in the standard hybrid model approach, treating every parton as independent from the moment when it is created at a splitting 
--- without regard for resolution. 
We then step through time incrementally, starting  
from when a parent splits into its offspring. At each time 
step we recalculate $\lres$ from the temperature at the parent's position, and we advance the offspring particles' positions according to their initial velocity, which the energy loss procedure (\ref{CR_rate}) from the hybrid model keeps constant.\footnote{This is true only when transverse broadening, 
described in Section~\ref{HybSum} and parametrized by $K$, is absent.  If $K\neq0$, the resulting
kicks to the transverse momentum of the
partons in the shower would change their velocity vectors, and would introduce changes
to the separation between offspring partons as they propagate through the plasma.  
The calculations of Ref.~\cite{Casalderrey-Solana:2016jvj} indicate that including a nonzero $K$
has negligible effects on all the observables that we shall consider in this paper.
For this reason, as well as because including $K\neq 0$ would considerably complicate
our implementation of resolution effects, we set $K=0$ throughout this paper.}  
Note also that we make no changes to the shower before
$\tau=0.6$~fm$/c$ since before this time we have no hydrodynamic medium and no parton energy loss in the model. In effect, we 
take $\lres=\infty$ before $\tau=0.6$~fm$/c$ and choose $\lres=\rres/(\pi T)$ at $\tau=0.6$~fm$/c$, 
and from then onward.
Starting at $\tau=0.6$~fm$/c$ and at each time-step thereafter, we ask whether each set of sibling partons is resolved or unresolved, and if they are unresolved we group them into an effective parton.
We record the first time-step at which  the separation between a pair of offspring exceeds $\lres$, with $\lres$ defined in terms of the temperature at the location of their parent effective parton.
This is the time after which each offspring 
will begin to lose energy independently.  
We also record the position of each of the offspring at this time.
After applying this procedure throughout, we record the modified lifetimes (from the time when a parton first begins to lose energy independently at its own resolution time until the time when it is replaced by its
two independent offspring at their resolution time) for each of the particles in the shower.

At each time step, we require that each parent in the branching tree 
must be resolved if any of its offspring have been resolved.  If a pair of offspring 
are resolved before their parent, we set the resolution time of the parent to that
of the offspring, in effect setting the modified lifetime of the parent to zero. (For an example, consider 
$s_3$ in the lower panel of Fig.~\ref{Fig:tree}.)
At each time step, we also enforce that sibling particles must resolve at the same time, so that if one sibling's resolution is forced by one of its children or grandchildren, then the other sibling's is as well; if a violation is found, we set the later resolution time equal to the earlier. 
We keep reapplying these two checks in alternation until doing so does not change anything, and only then proceed.
With the continued application of these two checks, when offspring particles resolve they force the resolution of their parent particle, and their parent particle's parent, and so on. Thus, sibling particles very late in the event which separate very rapidly from each other can drastically change the resolution time of every particle in the siblings' lineage.

After stepping through the entire history of the branching tree, applying the algorithm that
 we have just described, we end up with a new tree, with new creation times and positions for offspring particles (at the moment when they were resolved), and new lifetimes for each particle in the tree.  We then quench this new shower, applying the energy loss rate
 (\ref{CR_rate}) to the partons (effective partons) in this new shower using their new positions, creation times, and lifetimes but otherwise following the hybrid model algorithm as described in
 Section~\ref{HybSum} and Refs.~\cite{Casalderrey-Solana:2014bpa,Casalderrey-Solana:2015vaa,Casalderrey-Solana:2016jvj}. 

Now that we have explained the algorithm via which we shall model the effects
of resolution in the hybrid model, an algorithm that has the virtue of being fully specified with
no remaining unresolved ambiguities but that is surely not unique, we are ready
to look at how turning on $\lres\neq 0$ modifies the predictions of the model for
some important jet observables.

\section{\label{Effect}The Effects of Resolution on Jet Observables}

As already outlined in Section~\ref{HybSum}, the analysis of 
resolution effects in this work is built on top of the hybrid model without transverse broadening effects, but with particles originating from the backreaction of the medium --- the wake that the jet creates in the plasma --- included. 
Results for observables are obtained from hadrons that come from the hadronization
of the parton showers originating in hard collisions as well as those that come from the
hadronization of the medium, including in particular the perturbations originating from the wake.
In order to compare to data, before reconstructing jets we need to perform a background subtraction, using the same
techniques for doing so that the experimentalists have used in their analysis of whichever
observable we wish to compare to.  
The details of these procedures are described in Ref.~\cite{Casalderrey-Solana:2016jvj}.

\begin{figure}[t]
	\centering
	\includegraphics[width=.87\textwidth]{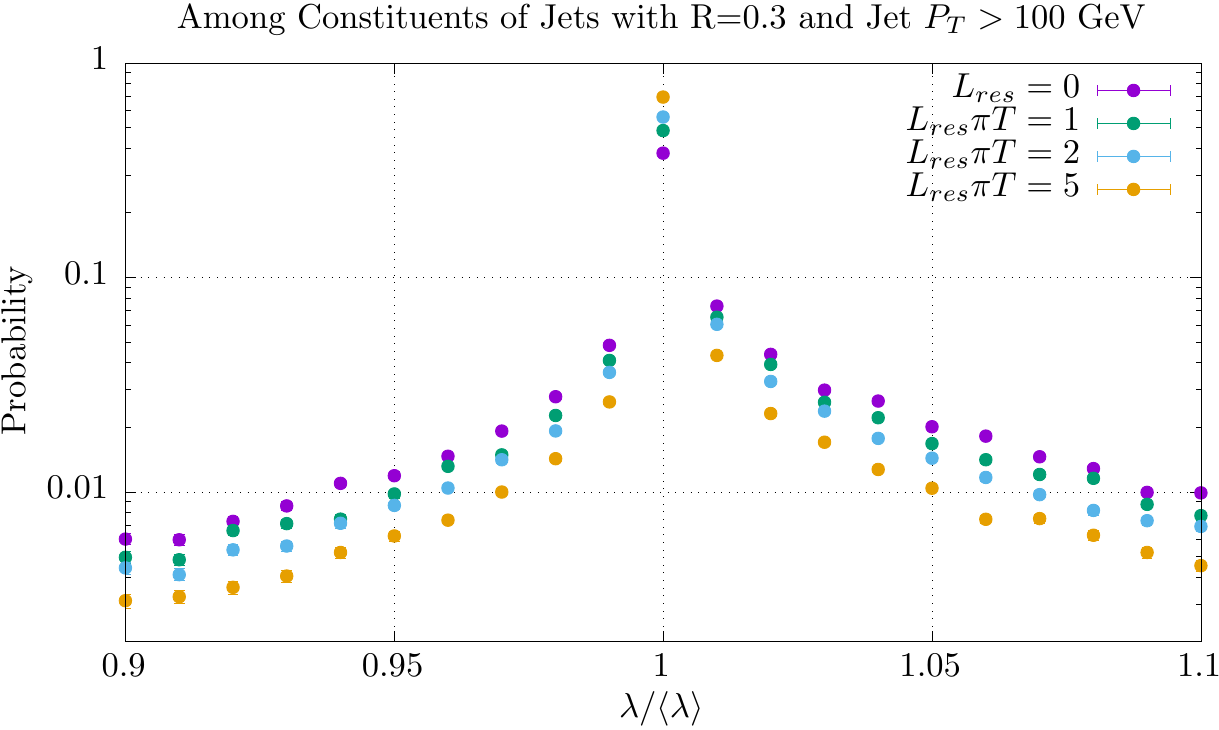}
	
	\caption{\label{Fig:lambda} 
	To illustrate the effects of resolution, we plot the distribution of $\lambda/\langle \lambda \rangle$ averaged over a sample of jets with $p_T>100$~GeV from 0-5\% centrality collisions with 
$\sqrt{s_{NN}}=2.76$~TeV.	We have quenched the partons in the jet showers using the
hybrid model with varying values of $\lres$, in all cases with the fitted value of $\aSC$.
$\lambda$ is the ratio of the energy of a particular parton after quenching to the value it would have had in the absence of the medium; $\langle \lambda \rangle$ is the average of $\lambda$
	over all the partons in a single jet.
As the resolution length $\lres$ increases, more and more of the partons within any given jet remain unresolved, behaving as a single effective parton and losing the same fractional energy.
Hence, as $\lres$ is increased this distribution tends toward a $\delta$ function at 1. (Even as $\lres\to\infty$
it does not become a true $\delta$ function, though, because the partons reconstructed as a jet
are not all descendants of a single parent parton.)
 }
\end{figure}

\subsection{Distribution of energy lost by partons in a jet}

Before turning to actual jet observables, we want to begin by getting some insight into how increasing $\lres$ affects the distribution of the amount of energy lost by the set of partons in
the final state of the parton shower (i.e.~the on-shell partons that are ready to hadronize) within
a high energy jet.  
For each such final parton coming from the parton shower we define the quantity $\lambda\equiv E_Q/E_V$, where $E_V$ is the energy that this particular parton would have had had in vacuum, in the absence of any medium, and $E_Q$ is the quenched energy that this particular parton has after it has propagated through the medium. 
(This is a quantity that we can look at within our model calculation. It
cannot be determined in experimental data, of course, because experimentalists do
not see partons, because even if they did they would have no way of knowing whether a final parton came from the parton shower, and because even if they did they would have no way of knowing $E_V$.)
Now, if all the partons within a jet had lost the same fraction of their energy, which
is to say if all of them had behaved throughout as a single effective parton as would
be the case for $\lres\to \infty$, then
every final parton in the shower would have the same $\lambda$.
Therefore, one way to see that because $\lres$ is finite 
some partons within the shower have indeed been resolved, and
treated independently by the medium, is to compute, jet-by-jet,
the distribution of the quantity $\lambda/\langle \lambda \rangle$.
Here, $\langle\lambda\rangle$ is the average of $\lambda$ for all the partons in a single jet.
We can then compute the $\lambda/\langle \lambda \rangle$ distribution
averaged over many jets. 
 We show the result in Fig.~\ref{Fig:lambda} for jets with $p_T>100$~GeV 
 reconstructed by applying the anti-$k_t$ algorithm with $R=0.3$ to the partonic final state from our hybrid model with four different
 values of $\rres \equiv \lres \pi T$, namely 0, 1, 2 and 5. 
 As expected, the larger the value of $\rres$ the more this distribution peaks 
 around $\lambda/\langle \lambda \rangle=1$, getting closest to a $\delta$ function distribution for the case $\rres=5$. 
 Note that not all of the partons that are reconstructed as part of a particular jet
 come from a single parent parton from the hard scattering; some come from initial state
 radiation, or even from a different parton from the hard scattering.
 For this reason, even when $\rres$ is as large as 5 the distribution has significant tails.

The reader will observe that in the opposite limit, when $\lres=0$ and the medium perfectly
resolves the partons in the shower as in the unmodified hybrid model, treating each parton as losing energy independently from the instant that it is created in a splitting,
the distribution
in Fig.~\ref{Fig:lambda} is already fairly peaked around 1.  There are two reasons for this.
First, the partons that lose all their energy to the medium and do not make it
into the final state of the quenched shower are excluded from this plot; 
only those that survive, and hence can be reconstructed as belonging to a jet, can contribute.
Many partons that were resolved by the medium and have a $\lambda$ well below $\langle\lambda\rangle$ therefore do not end up counted in the histogram plotted in Fig.~\ref{Fig:lambda}.
The second reason arises from the bias imposed by looking only at reconstructed jets
with $p_T$ above some cut, here 100 GeV.  This cut, together with the fact that
the jet spectrum falls steeply with $p_T$, biases the sample of jets toward those whose
initial energy was not far above 100 GeV and which did not lose much energy.
Since wider jets~\cite{Chesler:2015nqz,Rajagopal:2016uip} 
containing more, and softer, partons lose more energy than narrower jets
containing fewer harder partons~\cite{Milhano:2015mng,Casalderrey-Solana:2016jvj}, this selection criterion biases the sample toward narrow
jets containing fewer harder partons 
which are less likely to separate from each other sufficiently
to be resolved and which travel longer distances before they themselves split.

\subsection{Jet $R_{AA}$, including its Dependence on the Jet Reconstruction Parameter $R$}

Turning now to experimental observables, henceforth in all analyses that we report we include particles coming from
the medium perturbed by the wake left in it by the jet as well as particles coming 
from the parton shower. Note that for all observables that we look at in this
paper, we will focus on Pb-Pb collisions with $\sqrt{s_{NN}}=2.76$~TeV.

The first observable that we must calculate is the jet $R_{AA}$, because we use the 
experimental measurement of this observable to fix $\aSC$, the parameter that controls 
the magnitude of the energy loss in the hybrid model, see  Eq.~(\ref{CR_rate}). 
As we noted in Section~\ref{HybSum}, in the hybrid model we fix $\aSC$ by matching the predictions
of the hybrid model to the CMS measurement
of jet $R_{AA}$ for jets reconstructed with the anti-$k_t$ reconstruction parameter $R=0.3$
that have  100~GeV$<p_T<$~110 GeV in the 10\% most central Pb-Pb collisions~\cite{Raajet:HIN,Khachatryan:2016jfl}.
We obtain four different values of $\aSC$ by matching the predictions of the
hybrid model for this one data point to the upper and lower ends of the CMS error bar
and doing so with the temperature at which we turn quenching off set to 145~MeV and 170~MeV.
As in Refs.~\cite{Casalderrey-Solana:2014bpa, Casalderrey-Solana:2016jvj, Casalderrey-Solana:2015vaa}, 
for any observable that we look at later, we calculate the predictions of the
hybrid model using the highest and the lowest of these four values of $\aSC$, in so doing
obtaining a band for the hybrid model prediction 
that encompasses the experimental error in the measured data point
that we use to fix $\aSC$ plus a crude estimate of what can loosely be considered
a theoretical systematic error coming from the model~\cite{Casalderrey-Solana:2014bpa, Casalderrey-Solana:2015vaa}.

Here, we must fix  the fitted range for $\aSC$ anew for each nonzero value of $\lres$ that we employ, namely $\lres=\rres/\pi T$ with $\rres=$~1, 2, and 5.  
Without doing the calculation, it is not obvious whether the fitted values of $\aSC$
should increase or decrease with increasing $\rres$.  Increasing $\rres$ means
that various parent partons live longer, and can lose fractionally more energy, since in the expression (\ref{CR_rate}) for $dE/dx$ the rate of fractional energy loss increases with increasing
$x/x_{\rm therm}$.  But, it also
means that twice as many offspring partons live less long, and hence lose less energy.
If the first effect were to dominate, $\aSC$ would need to be reduced as $\rres$ is
increased, in order to maintain the fit to the experimentally measured jet $R_{AA}$ data point,
with its error bar.  In fact, what we find instead is that the fitted range for $\aSC$
increases modestly with $\rres$, increasing relative to its $\rres=0$ range of 0.323~$<\aSC<$~0.421 by about 6-7\%, 9-10\% and 13-16\% for $\rres=$~1, 2 and 5.
The fact that the fitted range of $\aSC$ increases slightly with increasing $\lres$ indicates that
the consequence of including a nonzero resolution length $\lres$ that impacts jet energy loss most is the fact that various offspring partons have shorter lifetimes and so lose less
energy.  In order to fit the jet $R_{AA}$ experimental data 
point, the reduction in the number of effective partons caused by delaying the appearance of
offspring partons as effective partons
until they have been resolved and thus shortening the lifetime of
the offspring partons as effective partons losing energy independently) needs to be compensated by increasing $\aSC$ so as to shorten the thermalization distance (\ref{CR_xstop}) in the hybrid model, increasing the rate of energy loss for all partons.

\begin{figure}[t]
	\includegraphics[width=.5\textwidth]{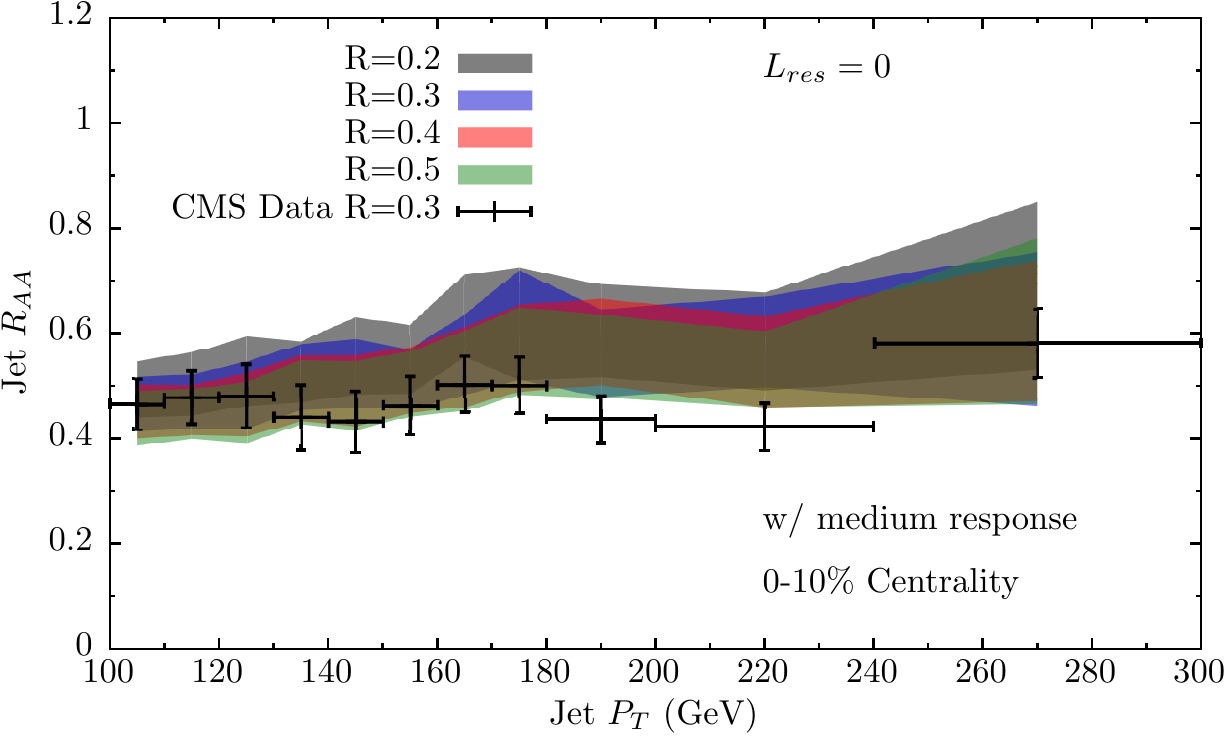}
	\includegraphics[width=.5\textwidth]{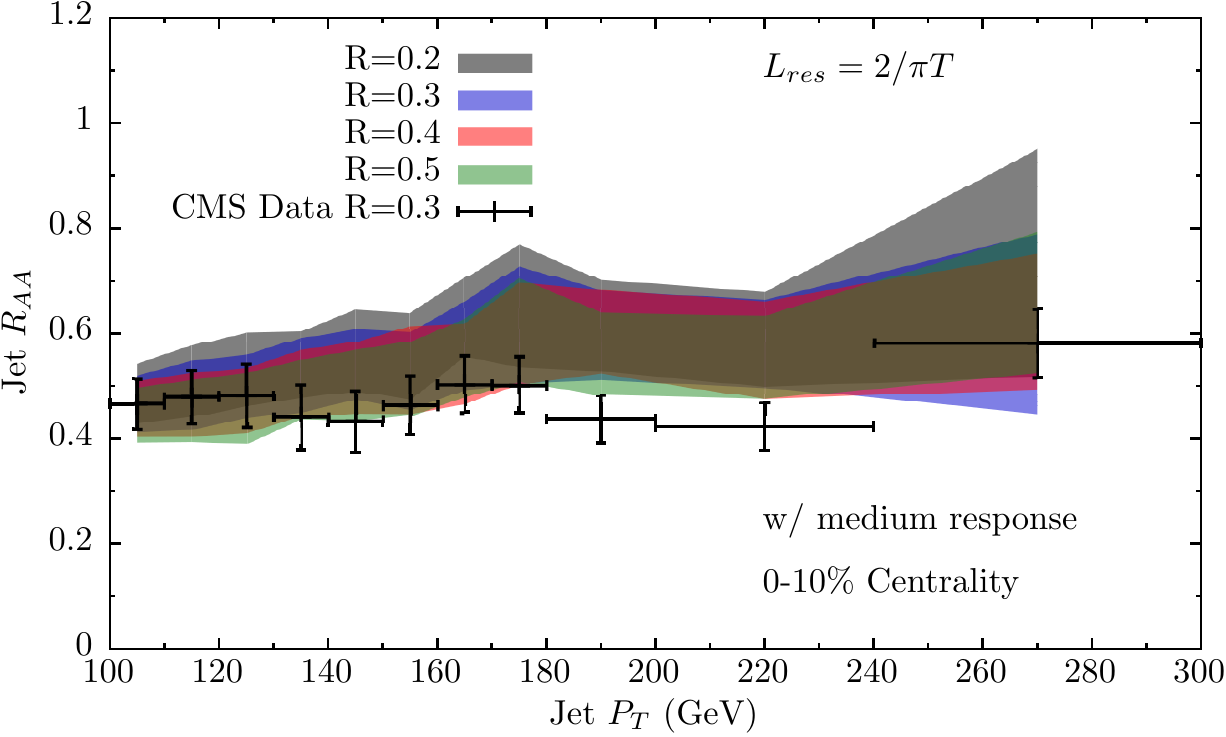}
	\caption{\label{Fig:RAA} Jet $R_{AA}$ as a function of jet $p_T$ for various values of the  reconstruction parameter, $R$, for $\lres=0$ (left panel) and $\lres=2/\pi T$ (right panel).
We see that wider jets tend to lose more energy than narrower jets. Including the
effects of resolution does not affect this conclusion.
	}
\end{figure}

With $\aSC$ fixed, the simplest observable for us to compute is the jet $R_{AA}$, but now using
varying values of the anti-$k_t$ reconstruction parameter, $R$.  
Reconstructing a jet sample using a smaller value of $R$ yields a sample in which 
narrower jets dominate, for two reasons: nearby clusters are more likely to be
counted as separate jets, and the full energy of a wider jet may not be reconstructed,
making it less likely for wider jets to pass the $p_T$ cut. 
We show our results in Fig.~\ref{Fig:RAA}, with $\lres=0$ and with $\lres=2/\pi T$.
In both panels,  the blue band agrees with the left-most CMS data point
because we have used this point to fit $\aSC$.
We see that wider jets tend to lose more energy than narrower jets, resulting in more suppression of their $R_{AA}$. (One may speculate that increasing $R$ and reconstructing wider jets could mean catching more of the ``lost'' energy within the reconstructed jets,
which would mean less suppression of jet $R_{AA}$.  Clearly this effect does not dominate,
at least up to $R=0.5$.  Much of the ``lost'' energy ends up at larger angles relative to the jet axis.)
The result that wider jets tend to lose more energy than narrower jets in the hybrid model was
already noted in Ref.~\cite{Casalderrey-Solana:2016jvj}; here we see that this conclusion
remains unchanged, and in fact the hybrid model results for $R_{AA}$ are hardly modified, when
we include resolution effects --- as long as we refit $\aSC$.
This is an indication of the robustness of the hybrid model, including in particular the procedure of fitting the
single parameter that controls the rate of energy loss to an experimentally measured jet $R_{AA}$
data point.
(The conclusion that wider jets lose more energy also
arises for holographic jets~\cite{Chesler:2015nqz,Rajagopal:2016uip};
and, the conclusion that jets containing more effective
partons lose more energy also arises 
at weak coupling~\cite{Milhano:2015mng,Escobedo:2016jbm}.)
It is also worth noting that the $R$-dependence of jet $R_{AA}$ that we find --- namely slightly less suppression of $R_{AA}$ for the narrower jets  reconstructed using smaller values of $R$ --- is similar to what has been seen in recent measurements from CMS~\cite{Khachatryan:2016jfl}, although
at present the error bars are too large relative to the smallness of the $R$-dependence to allow
for a definitive statement.
Increased precision for this type of measurement is important as it could yield further confirmation that
narrower jets lose less energy, and that the lost energy seems to efficiently end up at
large angles relative to the jet direction, for example as in a strongly coupled picture in which the
lost energy ends up in a hydrodynamic wake in the plasma.

\subsection{Fragmentation functions and jet shapes}

We have seen in the previous two Subsections that the important consequence of
introducing the effects of resolution comes
from the fact that the time when the medium notices that a parent 
parton has split into two offspring is delayed until the offspring partons
can be resolved, meaning that partons with moderate energies have
less time to lose energy independently. Because they spend more time
losing energy as a part of their larger parental effective parton before
becoming independent, they are more likely
to lose only the same fraction of their energy as the jet as a whole, not more (Section 5.1).
As a second consequence, the total energy lost by the jet as a whole is reduced, but
this is compensated for in the hybrid model by increasing $\aSC$, the parameter that
controls the magnitude of parton energy loss (Section 5.2) in order to maintain
the fitted agreement between the hybrid model prediction for jet $R_{AA}$ and data.
Both because of this compensation via increasing $\aSC$ and because they survive
longer as effective partons, the hardest, most energetic, partons in the core of
the jet suffer a modest increase in their energy loss.

The modification of the 
spacetime structure of the jet parton shower, as perceived by, modified by, and responded
to by, the strongly coupled liquid QGP in which it finds itself resulting from 
introducing the effects of resolution should manifest itself
in modifications to various intra-jet observables, to which we now turn.
Based on what we have seen in Sections 5.1 and 5.2, we  expect that, relative
to previous hybrid model studies in which $\lres=0$, once we turn on a nonzero $\lres$ 
the moderate energy partons in the jet should lose somewhat less energy and
should be more likely to survive into the final state, where they can hadronize and be
seen as moderate energy hadrons within the reconstructed jets.  Correspondingly, the few hardest partons and hence hadrons should
lose a little bit more energy.  After hadronization, we should expect to see both effects in the 
fragmentation function.  Also, to the extent that the few hardest partons define the center of the jet and
the moderate energy partons are more spread out in angle around that center, we should
also expect to see both effects in the jet shape.
We therefore look at how the hybrid model predictions for 
these two intra-jet observables are modified once we turn on a nonzero $\lres$.

\begin{figure}[t]
	\includegraphics[width=.5\textwidth]{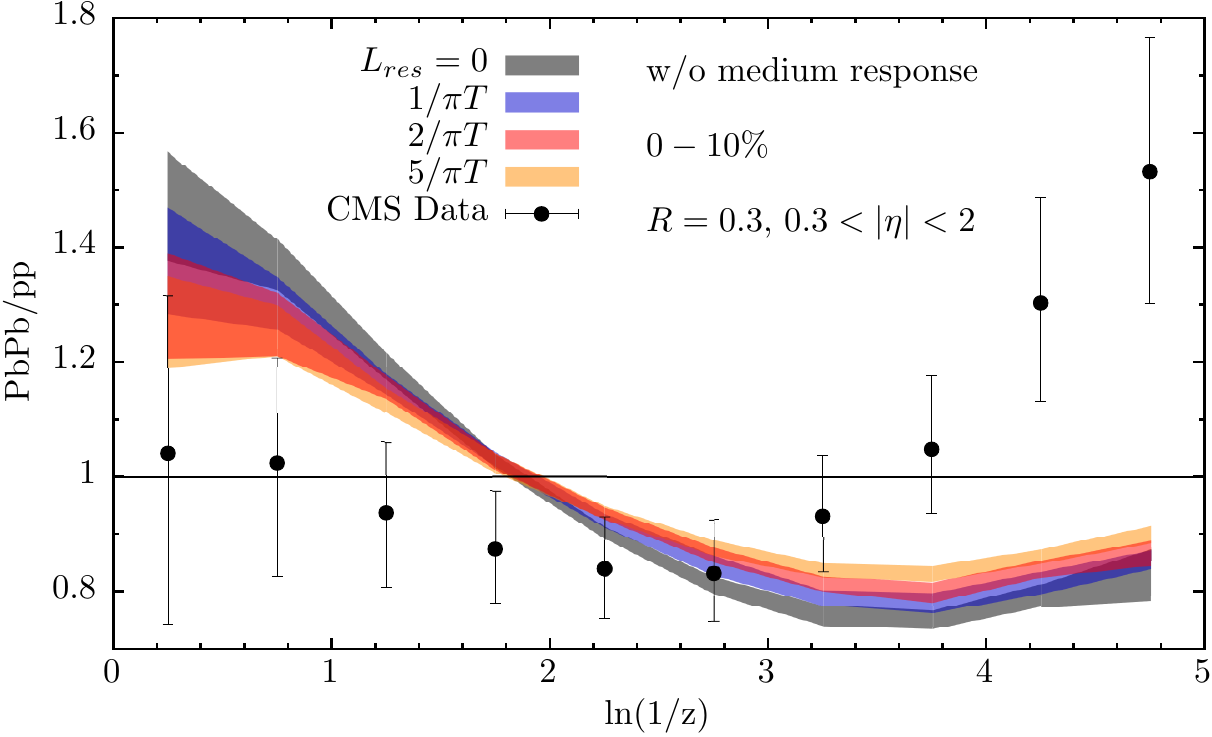}
	\includegraphics[width=.5\textwidth]{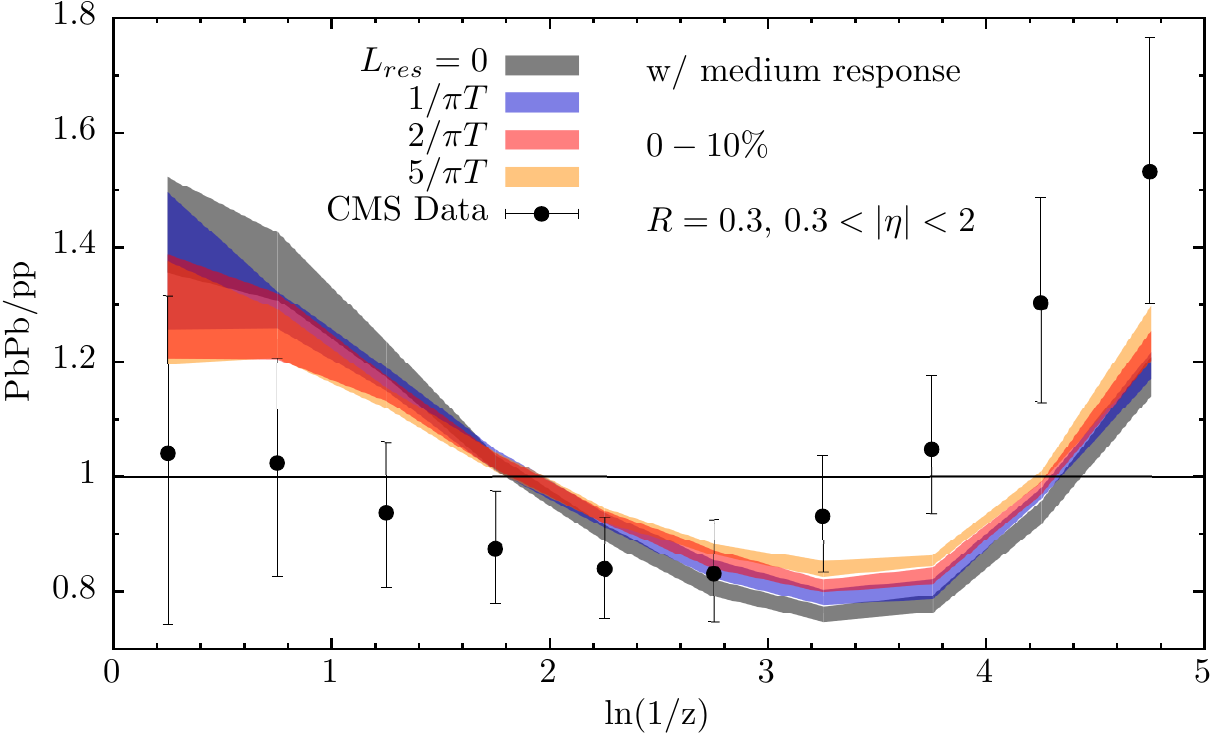}
	\caption{\label{Fig:FF} Fragmentation functions characterize the probability distribution for the longitudinal energy fraction $z$ carried by an individual hadron relative to the total energy of the jet. 
	We plot the ratio of the fragmentation function for jets reconstructed
	with the anti-$k_t$ parameter $R=0.3$ that have $p_T^{\rm jet}>100$~GeV 
	and $0.3<|\eta|<2$
	in the 10\% most central Pb-Pb collisions to
	that in p-p collisions with the same 2.76 TeV collision energy.
	In the left panel, we only include hadrons coming from the hadronization of the jet showers.  In the right panel, we also include hadrons coming from the medium after background subtraction, meaning that we see the effects of the wake that the jet leaves behind in the plasma. 
	In both panels, we show the predictions of the hybrid model with $\rres=\lres\pi T$ given
	by 0, 1, 2 and 5.  Turning on a nonzero $\lres$ has allowed more hadrons carrying 
	a smaller fraction of the jet energy to survive into the final state, seen on the right of each panel,
	and has correspondingly reduced the contribution of hadrons carrying a large fraction of the jet energy, seen on the left.  
	Including the effects of resolution shifts the predictions of the hybrid model in the direction of the CMS data~\cite{Chatrchyan:2014ava}, but the effects are relatively small in magnitude, in particular for
	$\rres=1$ and 2 which corresponds to our range of expectations for the resolution length $\lres$.
}
\end{figure}

We begin with the jet fragmentation function, 
which characterizes the fraction of the longitudinal energy of the jet carried by individual
charged hadrons within a jet.
The tracks used in the analysis lie within a distance $r<R=0.3$ from the jet axis 
in the $\eta-\phi$ plane, and the distribution 
is expressed in terms of the variable $\rm{ln}(1/z)$, with $z$ defined 
as $z \equiv p_{\rm track} \cos \theta /p_{\rm jet}$, where $\theta$ is the angle between 
the momentum vector of the track and the jet axis as defined via the anti-$k_t$ reconstruction.
In Fig.~\ref{Fig:FF}, we show the ratio of the quenched fragmentation function for jets
in heavy ion collisions to the
fragmentation function for jets produced in p-p collisions that propagate in vacuum.
In the right panel, we include the effects of the wake in the strongly coupled plasma
that the jet creates since, as described in Section~\ref{HybSum}, some of the
hadrons that come from the hadronization of the plasma including this wake must end
up included in the reconstructed jet.  In the left panel, we do not include this response 
of the medium to the jet. 
In both panels, the previous predictions of the hybrid model
with $\lres=0$ are shown in grey, whereas our results for the physically well-motivated
values of the resolution length, $\lres=1/\pi T$, and
$\lres=2/\pi T$ are shown in blue and red.  The orange band corresponds to
the unphysically large value $\lres=5/\pi T$.
We see in Fig.~\ref{Fig:FF} that, as expected,
the contribution from the hardest tracks lying around $z\lesssim 1$ is diminished 
by increasing $\lres$, while the contribution of 
the softer particles with moderate energies around $\rm{ln}(1/z)\sim 3$ is enhanced. 
Including the effects of resolution shifts the predictions of the hybrid model in the direction 
of the CMS data~\cite{Chatrchyan:2014ava}, 
but the rise seen in the CMS data at the smallest $z$ (largest $\rm{ln}(1/z)$) 
is not fully explained.  We also see that in this regime the contribution coming from including
the backreaction of the medium, the wake in the plasma, is larger in magnitude
than the contribution coming from including the effects of resolution.
For the hardest hadrons with $z\lesssim 1$, on the left, the two effects both push
the predictions of the model downward and together bring them quite close to the CMS data.

\begin{figure}[t]
	\includegraphics[width=.5\textwidth]{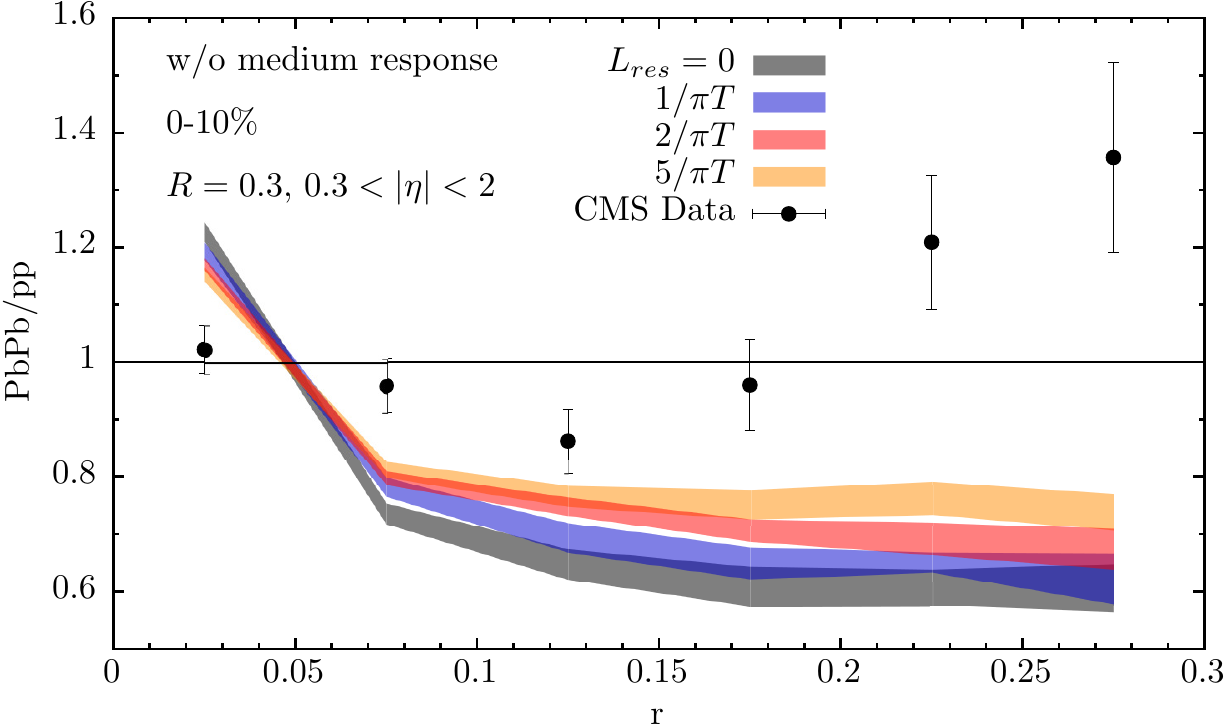}
	\includegraphics[width=.5\textwidth]{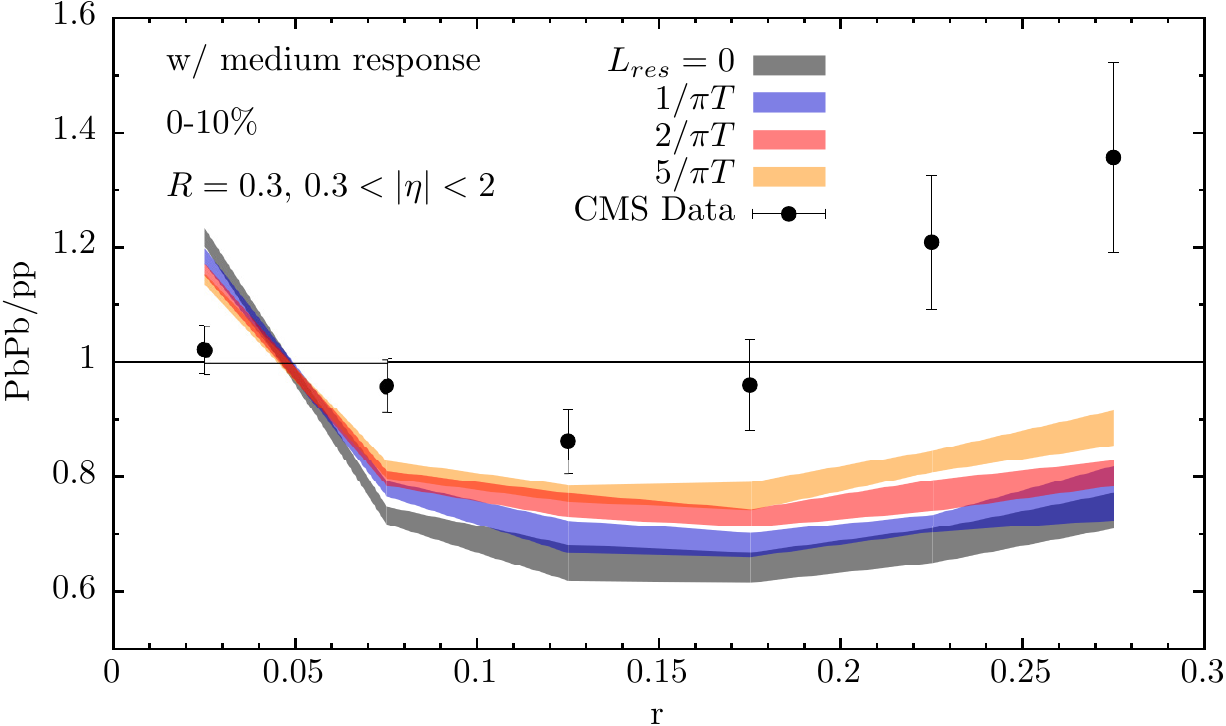}
	\caption{\label{Fig:Shapes} The jet shape observable characterizes the angular distribution of energy within the jet as a function of $r$, the angle in the $\eta-\phi$ plane relative to the jet axis. 
	We plot the ratio of the jet shape for jets reconstructed
	with the anti-$k_t$ parameter $R=0.3$ whose centers 
	lie within $0.3<|\eta|<2$ and
	that have $p_T^{\rm jet}>100$~GeV in the 10\% most central Pb-Pb collisions to
	that in p-p collisions with the same 2.76 TeV collision energy.
	In the left panel, we only include hadrons coming from the hadronization of the jet showers.  In the right panel, we also include hadrons coming from the medium after background subtraction, seeing the effects of the wake in the plasma. 
	In both panels, we show the predictions of the hybrid model with $\rres=\lres\pi T$ given
	by 0, 1, 2 and 5.  Turning on a nonzero $\lres$ has allowed more hadrons at larger angles relative to the jet axis to survive into the final state, see the right of each panel,
	and has correspondingly reduced the contribution of hadrons at the very center of the jet, 
	see the left of each panel.  
	Including the effects of resolution shifts the predictions of the hybrid model in the direction of the CMS data~\cite{Chatrchyan:2013kwa}, but the effects are relatively small in magnitude, in particular for
	$\rres=1$ and 2 which corresponds to our range of expectations for the resolution length $\lres$.
}
\end{figure}

We turn next to the jet shape observable, which quantifies 
the fraction of the total energy of jets reconstructed with anti-$k_t$ parameter $R$ that lies 
within an annulus of radius $r$, and width $\delta r$ (in $\eta-\phi$ space), centered on the jet axis. 
Following Ref.~\cite{Chatrchyan:2013kwa}, we define the differential jet shape as
\begin{equation}
\rho(r)\equiv \frac{1}{N_{\textrm{jets}}}\frac{1}{\delta r}\sum\limits_{\textrm{jets}} \frac{\sum\limits_{i \, \in \, r\pm\delta r/2}
	p_t^{i,\textrm{track}}}{p_t^{\textrm{jet}}} 
\label{eq:JetShapeDefn}
\end{equation} 
for $r<R$,
where the particles in the sum are all the hadrons found in the specified annulus (after background subtraction) whether or not they were identified as constituents of the jet by the anti-$k_t$ algorithm.
$\rho(r)$ is defined such that it is normalized to one. 
In Fig.~\ref{Fig:Shapes}, we show the ratio of the jet shape
for quenched jets in heavy ion collisions to that for jets produced in p-p collisions that
propagate in vacuum.
For reference, the experimental results for this ratio are also shown, as measured by the CMS collaboration~\cite{Chatrchyan:2013kwa}. 
In the right panel we include the effects of the wake in the strongly coupled plasma, and in
the left panel we do not include this response of the medium to the jet.
As in Fig.~\ref{Fig:FF}, in both panels of Fig.~\ref{Fig:Shapes} the colored bands show how the
predictions of the hybrid model change for $\rres=0$, 1, 2 and 5.
As expected, increasing $\lres$ increases the probability to find hadrons at
larger angles relative to the jet axis (and, as seen above, with moderate energies) making it into the detector and therefore into the jets.  
The energy fraction at the very core of the quenched
jets is depleted as a function of increasing $\lres$ and the contributions in wider
annuli are enhanced.
It remains the case, though, that because we are comparing quenched and unquenched
jets with the same final energy, because narrower jets lose less energy, and because
the jet spectrum falls rapidly with energy, there is a bias toward finding quenched
jets that are narrower than the unquenched jets. That is, the unquenched jets that were wider
lose more energy and end up below the $p_T^{\rm jet}$ cut used in the analysis, making
the jet shape after quenching narrower than that in p-p collisions.
As for the fragmentation function, including the effects of resolution shifts the predictions of the hybrid model for the jet shape seen in Fig.~\ref{Fig:Shapes} in the direction 
of the CMS data~\cite{Chatrchyan:2013kwa}, 
but the rise seen in the CMS data at larger angles is not fully explained.
We also see that in this regime  the contribution coming from including
the backreaction of the medium, the wake in the plasma, is comparable in magnitude
to the contribution coming from including the effects of resolution.

In the case of both the fragmentation function and the jet
shape, including the effects of resolution pushes the predictions
of the hybrid model in the direction of the data but the effects
are modest in magnitude meaning that the interesting qualitative
differences between the predictions of the model 
and the data that were noted in Ref.~\cite{Casalderrey-Solana:2016jvj} remain.

\subsection{Two missing-$p_T$ observables}

Ref.~\cite{Casalderrey-Solana:2016jvj} also provides hybrid model calculations
of a suite of intra-jet observables known collectively as missing-$p_T$ observables.
These characterize the distribution in momentum and angle of all the particles
in an event containing a pair of reconstructed jets, a dijet, 
with respect to the axis defined by the dijet~\cite{Khachatryan:2015lha}.
These are excellent observables with which to study the response of
the plasma to the jet: if one does not include the hadrons coming from the 
wake in the plasma
in the analysis, the predictions of the hybrid model for these observables
are in gross disagreement with experimental measurements~\cite{Casalderrey-Solana:2016jvj};
however, upon including the hadrons coming from the wake treated as in 
Ref.~\cite{Casalderrey-Solana:2016jvj} as summarized in Section~\ref{HybSum}, 
one obtains broad qualitative
agreement with the data while at the same time seeing very interesting remaining discrepancies.
The treatment of the wake assumes that the
energy and momentum deposited in the plasma equilibrates (more precisely,
hydrodynamizes) subject to energy and momentum conservation, and
that the resulting perturbation to the hydrodynamic flow and
the consequent perturbation to hadron spectra after hadronization are both
small enough to be treated to linear order.
The authors of Ref.~\cite{Casalderrey-Solana:2016jvj} speculate that
the interesting discrepancies between hybrid model predictions
and the experimental data could be ameliorated either by including
the effects of resolution or by improving the analysis of the wake, in particular
by taking into account the possibility that the wake does not
fully equilibrate.  After all, a part of the wake 
must be created not long before the jet exits the fluid and/or not long before
the fluid+wake hadronizes, meaning that some of the wake will have
very little time to hydrodynamize.
Here, we evaluate the effects of resolution and show that
they are small in magnitude relative to the discrepancies in question.  This provides
indirect support for the suggestion that further analysis of these discrepancies
will give us experimental access to the processes via which the wake --- a perturbation
to the hydrodynamic, strongly coupled, QGP --- relaxes by taking a snapshot of these
processes before they are complete.

The missing-$p_T$ observables are distributions of averages of the 
quantity $\mpt$, calculated for every track in the event and defined as 
\begin{equation}
\label{pt}
\mpt \equiv - \pt \cos\left(\phi_{\rm dijet}-\phi\right) \, ,
\end{equation} 
where $\pt$ and $\phi$ are the transverse momentum and the azimuthal angle of the track, respectively, and $\phi_{\rm dijet}$ is the direction defined by the dijet, namely the bisection of the 
azimuthal angle between the leading jet angle, $\phi_{\rm leading}$, and the flipped subleading jet azimuthal angle, $-\phi_{\rm subleading}$. 
With this definition, tracks in the subleading jet hemisphere give positive contributions to $\mpt$, while those in the leading jet hemisphere contribute negatively. 
If, as is natural, the hard partons in the subleading jet have lost more energy than those 
in the leading jet, that drives the average $\mpt$
of hard particles negative.  If, correspondingly, the
wake created by the subleading jet, with momentum in the same direction by momentum
conservation, is larger than the wake created by the leading jet, that drives the average $\mpt$ of 
soft hadrons coming from the wakes positive.
We consider dijet pairs 
reconstructed from hadrons lying within $|\eta|<2.4$ 
with leading and subleading transverse momenta $\pt^{\rm leading}>120$~GeV and $\pt^{\rm subleading}>50$~GeV respectively and with both jet axes within $|\eta|<2$. We also enforce a back-to-back criterion of $| \phi^{\rm leading}-\phi^{\rm subleading}|>5 \pi /6$ between the two jets. 
Although we initially consider all jets whose axes lie within $|\eta|<2$, we subsequently
make a further cut that restricts our sample of dijets to those for which both jet axes lie
within $|\eta|<0.6$.  (We use the larger pseudorapidity range initially to maximize the chance that we
do indeed find the two highest energy jets in the event; we use the smaller range for the analysis
in order to be able to look at hadrons out to relatively large angles away from the dijet axis.)
These are the same cuts used in the analysis of experimental data in Ref.~\cite{Khachatryan:2015lha}.

\begin{figure}[t]
	\centering
	\includegraphics[width=1.00\textwidth]{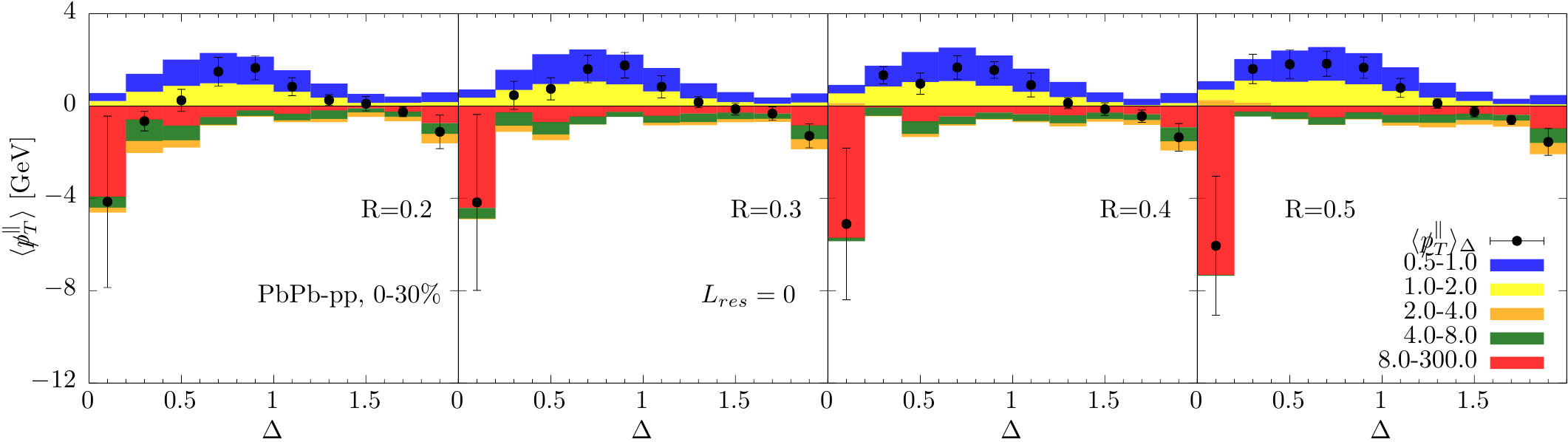}
	\includegraphics[width=1.00\textwidth]{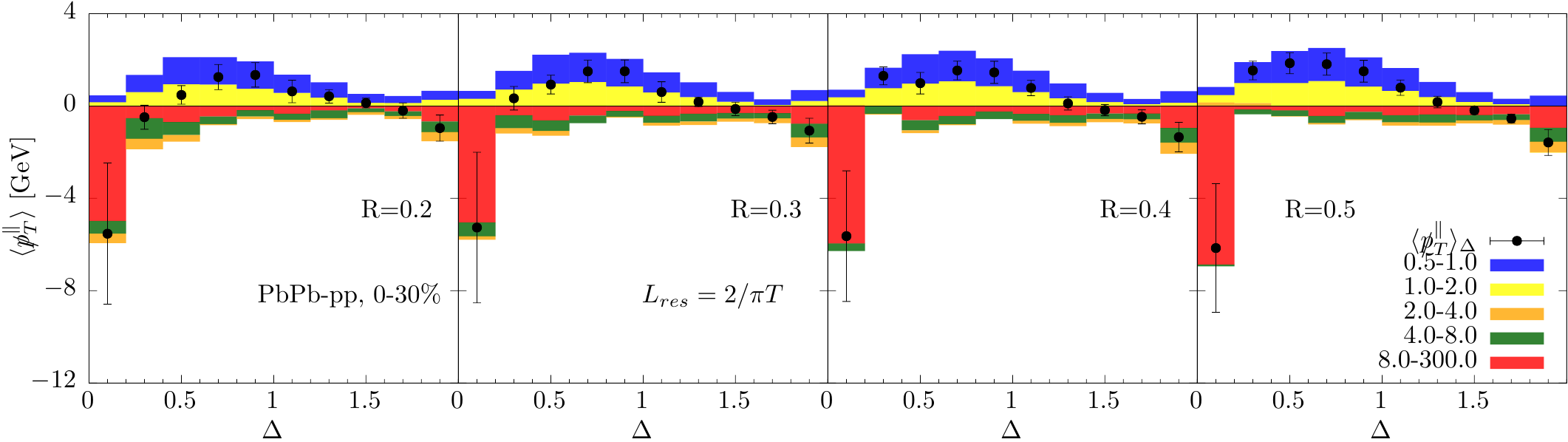}
	\includegraphics[width=1.00\textwidth]{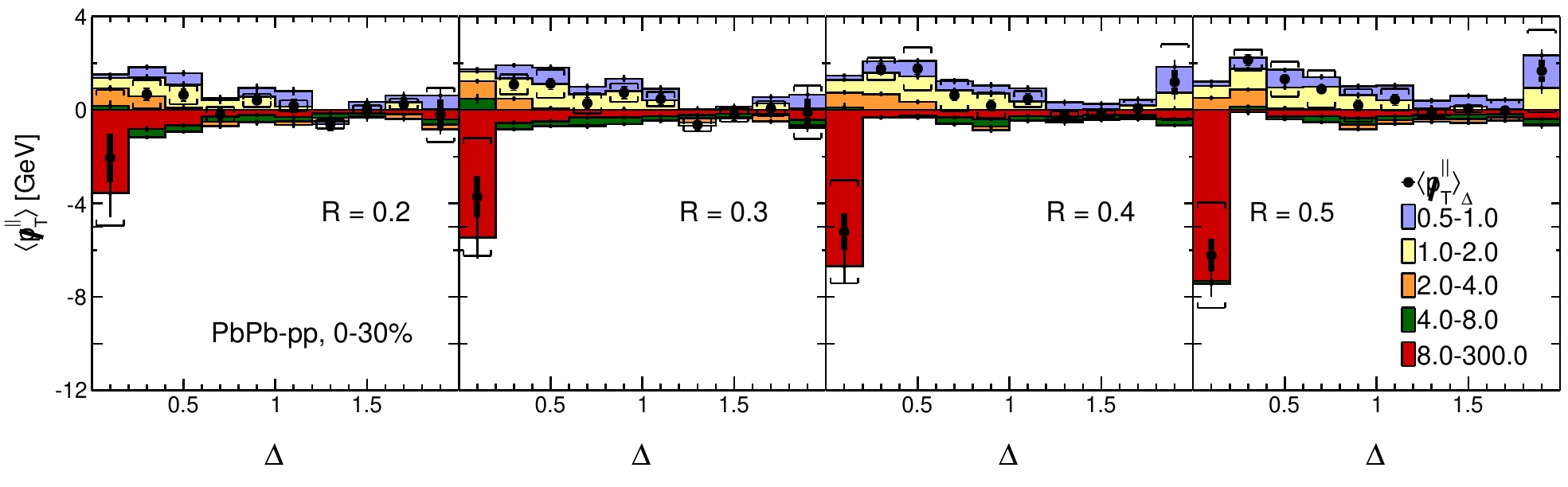}
	
	\vspace{-0.05in}
	\caption{\label{Rdependence} The observable $\langle \mpt\rangle$, 
	defined as the average of $\mpt$ in (\ref{pt}) over tracks,
	characterizes the momentum flow along the dijet direction with momentum pointing in the hemisphere of the leading/subleading jet counting negatively/positively and is referred to as the ``missing-$p_T$.'' Here, we plot the difference of the missing-$p_T$ in Pb-Pb and p-p collisions. The upper row of panels shows the predictions of the hybrid model
	with $\lres=0$ from Ref.~\cite{Casalderrey-Solana:2016jvj} for dijets reconstructed with
	four different values of the anti-$k_t$ reconstruction parameter, $R$.  
	The colored histograms represent the contributions
	to the missing-$p_T$ coming from hadrons in specified $p_T$ ranges, as a function of
	the angle relative to the dijet direction, $\Delta$; the points with error bars show the sum of the missing-$p_T$ for hadrons with any $p_T$.  
The lower row of
	panels shows experimental measurements of this observable from the CMS collaboration~\cite{Khachatryan:2015lha}. 	
	The middle row of panels shows our results, obtained from the hybrid model with $\lres=2/\pi T$. There are effects of resolution, namely differences between the middle and upper panels, but they are almost too small to be seen on the scale at which the plot is made, meaning that they
	are substantially smaller than the discrepancies between the hybrid model predictions in the upper panels and the experimental results in the lower panels.
		}
	\label{fig7} 
\end{figure}

In the upper and middle two rows of panels in Fig.~\ref{Rdependence}, 
we show our hybrid model results for the difference in the distribution of $\langle \mpt \rangle$, which is called the missing-$p_T$ and is the average of $\mpt$ over tracks,
in Pb-Pb versus p-p collisions 
as a function of $\Delta$ (the angular separation of the tracks in the $\eta-\phi$ plane with respect to either the leading or the subleading jet axis, depending on which yields a smaller $\Delta$) for dijet events reconstructed with different anti-$k_t$ radii $R$.
In the upper panels, $\lres=0$; these are the results of Ref.~\cite{Casalderrey-Solana:2016jvj}.
In the middle panels, $\lres=2/\pi T$; these are our results, including the effects of resolution.
The lower row of panels shows the experimental measurements
published by the CMS collaboration \cite{Khachatryan:2015lha}. 
The contributions to $\langle \mpt\rangle$ are further sliced into different $\pt$ bins, 
represented by the different colors, whose sum is shown by the black dots. 
Although there are effects of resolution in the middle panels of Fig.~\ref{Rdependence},
these effects, namely the differences between the middle and upper panels in the Figure, are almost too small to be visible.   This means that the effects of resolution seen in the jet shapes in Fig.~\ref{Fig:Shapes}, which were already small in magnitude, are reduced further by cancellation when we look at differences between leading and subleading jets as in the 
definition of the $\langle\mpt\rangle$ observable.

The particular discrepancy between the top and bottom panels of Fig.~\ref{Rdependence},
which is to say between the predictions of the hybrid model and experimental
measurements, that was highlighted in Ref.~\cite{Casalderrey-Solana:2016jvj} 
can be seen most clearly by looking at the orange histograms, namely the
contributions of hadrons with 2~GeV~$<p_T<4$~GeV.  In the hybrid model,
the orange contribution to $\langle\mpt\rangle$ is negative, meaning that in this $p_T$-range the greater
energy loss of the subleading jet is more important than the larger wake of the
subleading jet --- in the model.  In the experimental data in the bottom panels, 
the orange contribution to $\langle\mpt\rangle$ is positive, meaning that in reality the larger wake of
the subleading jet dominates in this $p_T$-range.  This means that
in the experimental data there are more hadrons coming from the wake with 
$p_T$ in this range than in the hybrid model.
Correspondingly, if we look at hadrons with $p_T<2$~GeV, the blue and
yellow histograms, we see that
there are fewer hadrons coming from the wake with $p_T$ in this softest
range in experimental data than in the hybrid model.
We now see, from the middle panels of Fig.~\ref{Rdependence}, that including
the effects of resolution does not significantly ameliorate these discrepancies.
This provides us with indirect evidence that these discrepancies are indeed
telling us about the inadequacies of our treatment of the wake and, in particular, are telling us
that the wakes that the jets leave in the plasma do not fully equilibrate, meaning that
the hadrons that come from these wakes are not as
soft in reality as in our fully equilibrated analysis.

\begin{figure}[t]
	\centering 
	\vspace{-0.2in}
	\includegraphics[width=0.64\textwidth]{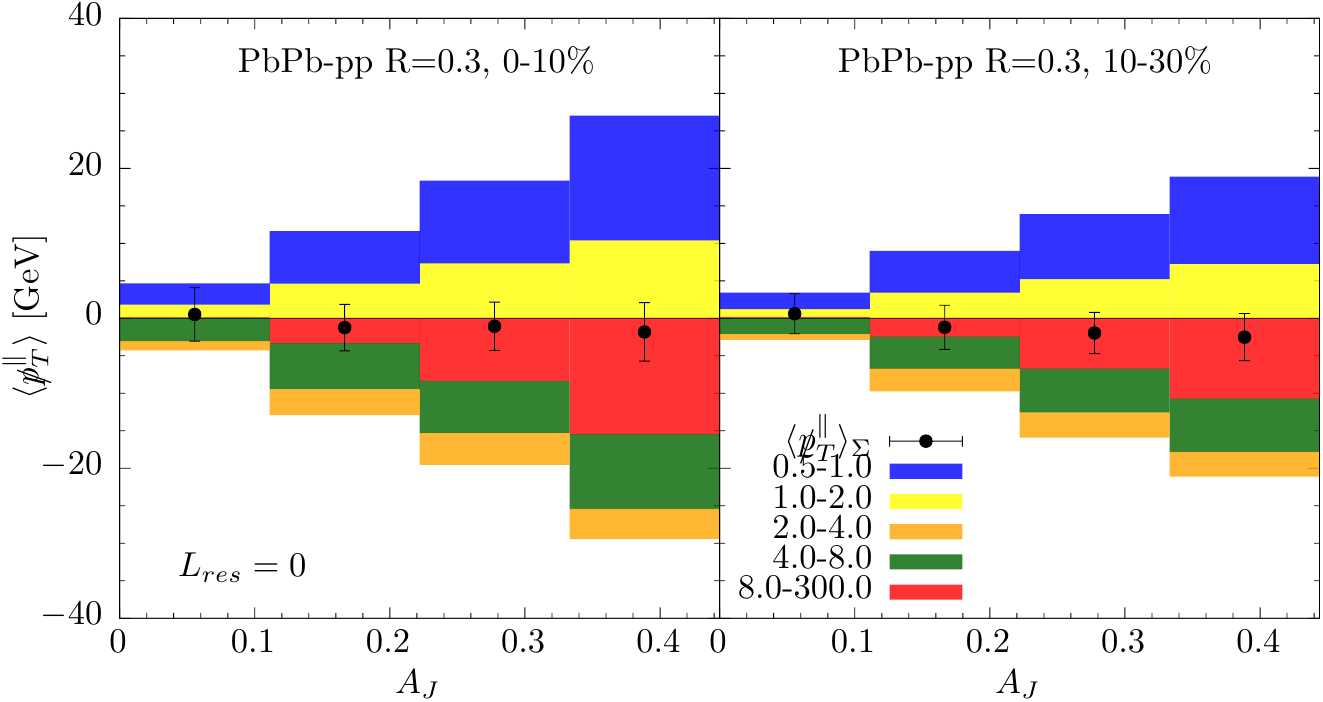}
	\vspace{-0.in}
	\includegraphics[width=0.64\textwidth]{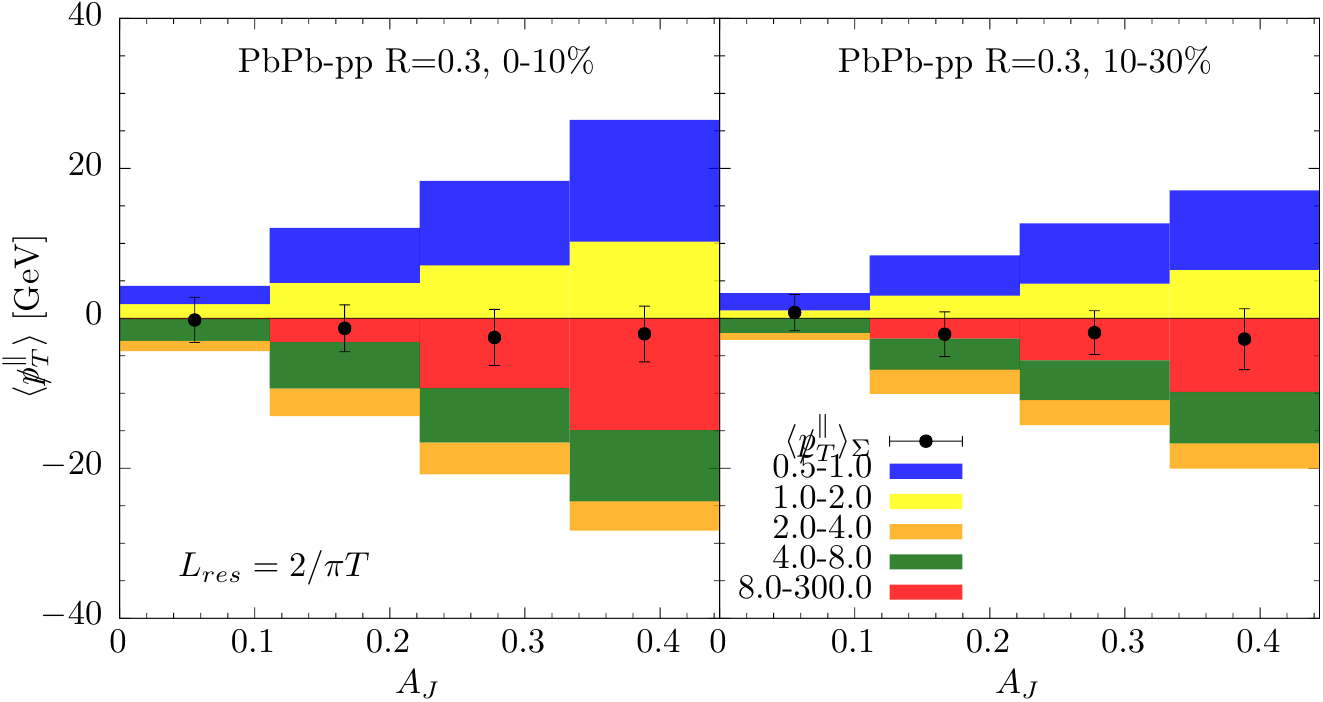}
	\includegraphics[width=0.67\textwidth]{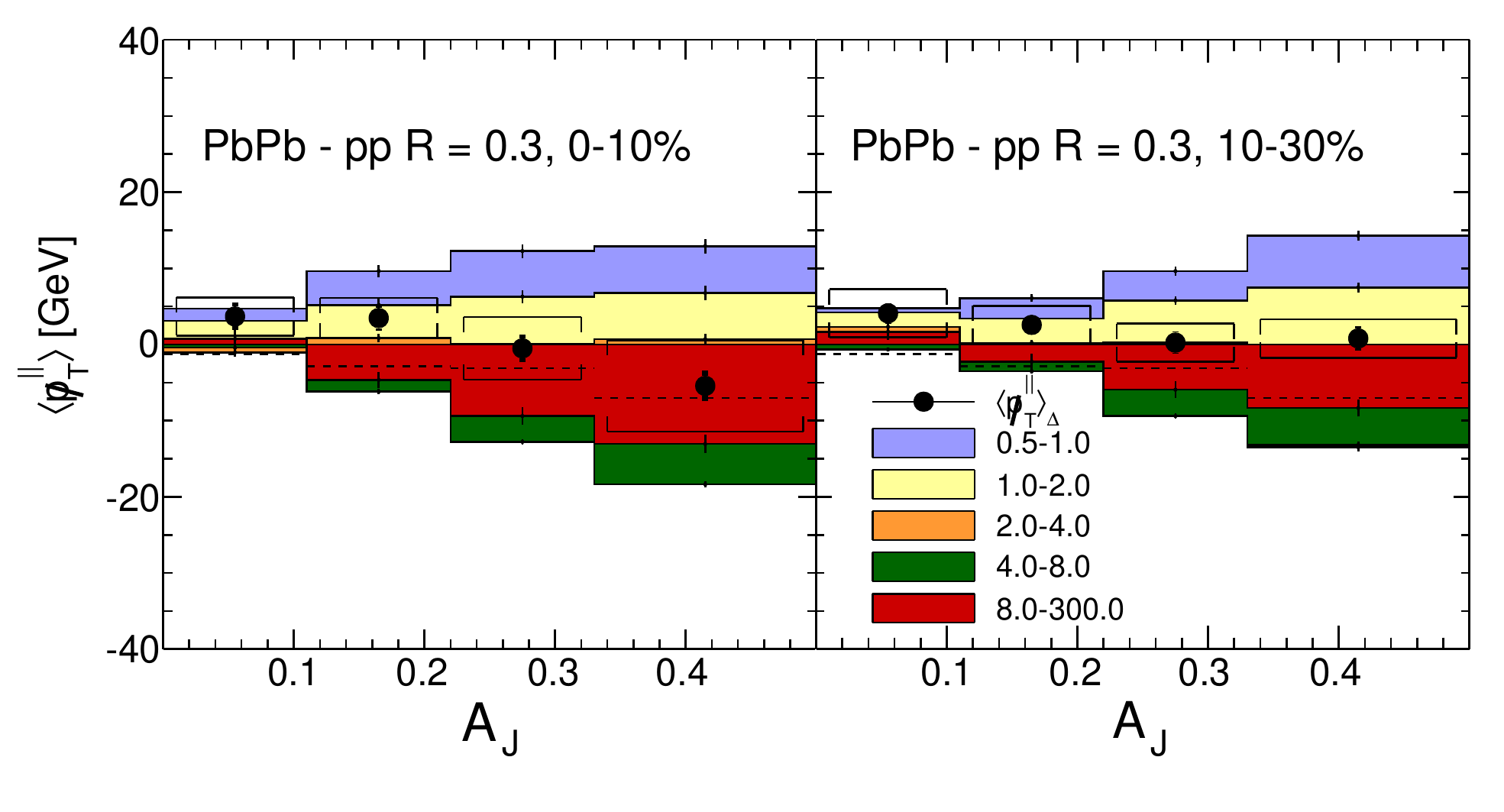}
	\vspace{-0.2in} 
	\caption{\label{Aj} This missing-$p_T$ observable characterizes the asymmetry between the spectrum of the leading and subleading jets, as in Fig.~\ref{Rdependence}, but this time integrated in $\Delta$ and sliced in terms of the dijet asymmetry variable, $A_J$.  The contributions of hadrons in different $p_T$-ranges to $\langle \mpt\rangle$ are color coded as before.
There are effects of resolution, namely differences between the middle and upper panels. However,
as in Fig.~\ref{Rdependence} they are almost too small to be seen, meaning that they
	are substantially smaller than the discrepancies between the hybrid model predictions 
	and the experimental results from Ref.~\cite{Khachatryan:2015lha} shown in the lower panels.
} 
\end{figure}

A different way to present the physics contained in Fig.~\ref{Rdependence} 
is to present the $\langle\mpt\rangle$ distributions 
integrated in $\Delta$, but sliced instead in terms of the dijet asymmetry 
variable $A_J$, defined as $A_J\equiv (\pt^{\rm leading}-\pt^{\rm subleading})/(\pt^{\rm leading}+\pt^{\rm subleading})$. This is done in Fig.~\ref{Aj}, for two different centrality classes in the left 
and right columns.
We only show dijets reconstructed with anti-k$_t$ $R=0.3$.
The top two rows of panels show the results of our simulations (with $\lres=0$ and $\lres=2/\pi T$)
and the bottom panels show CMS data~\cite{Khachatryan:2015lha}.  
Here again, the differences between the middle and upper panels, namely the
effects of resolution, are almost too small to be visible.  And, here again
these effects do not explain the discrepancies between the hybrid model
predictions in orange, as well as in blue plus yellow, and the experimental measurements.

\section{\label{Discuss}Discussion and Outlook}

In this paper, we have 
investigated the phenomenological implications for
various jet quenching observables of the fact that the strongly coupled QGP
produced in heavy ion collisions cannot resolve two partons in a jet shower
when they are closer together than some resolution length $\lres$.
Our study is exploratory in character, in three ways. First, it is built
upon the hybrid strong/weak coupling model for jet quenching, which
combines a weakly coupled treatment of jet production and showering
from \pythia with a treatment of parton energy loss that is patterned on
the rate of energy loss obtained via a holographic calculation in a different
strongly coupled gauge theory, as we described in Section~\ref{HybSum}.  
This model has been used successfully to describe various data sets
and gain various insights, but it {\it is} only a model, limited by its ingredients.
In many ways, the most interesting uses of a model like this are to find the
instances where it fails to agree with experimental data, as this 
teaches us much more than simply constraining the parameters of
the model can.  The analysis of this paper  provides an example.
Second, 
we assume that $\lres$ is of order the Debye length and hence choose $\lres=\rres/
\pi T$ with $\rres$ a constant that we treat as a new parameter of the model. As we described in
Section~\ref{Resin}, this is reasonable
for a strongly coupled plasma and in some, but not all, kinematical regimes in
a weakly coupled plasma.  We saw that in some kinematical regimes at weak coupling,
$\lres$ could be shorter than the Debye length which would reduce the observable
effects of resolution relative to what we have found.
We looked at results
for $\rres=1$ and 2, motivated by strong and weak coupling calculations of
the Debye length, as well as for $\rres=5$, which is a larger value than is physically motivated.
Third, even though it is inescapable that the plasma has some
nonzero resolution length $\lres$ and so must perceive the jet shower in terms
of a collection of effective partons,
our implementation of
the effects of resolution is  simplified considerably, as we described in Section~\ref{Implement}.
When a parent parton splits into two offspring, we treat the offspring as a single effective
parton with the energy, energy loss, and trajectory of the parent until the offspring are separated
by more than $\lres$ in the laboratory frame, and at that instant in time in the laboratory
frame we replace the single effective parton by the two offspring which henceforth begin
to lose energy independently.  
The need to specify a frame, together with certain other aspects of the way in which we build effective partons 
that we describe in Section~\ref{Implement}, has a certain arbitrariness to it
because we are not treating the (quantum mechanical)
processes via which resolution occurs in full. The virtue of our implementation
is that it constitutes a simple and unambiguous prescription that allows us to assess the
magnitude of the effects of resolution on various jet and intra-jet observables
upon exploring a set of reasonable values for the resolution length, $\lres$.
All three of these exploratory aspects of our study could be revisited in future
work, although the motivation for such efforts is reduced 
given that we find that including a reasonable
nonzero value for $\lres$ results in only modest changes to the model predictions for
the observables that we have investigated.

As we increase $\lres$, the number of effective sources of energy loss is reduced since
it takes longer for the medium to resolve the partons formed after each splitting. 
We saw in Section 5.1 that this increases the number of final state partons with moderate
energies whose energy
loss is the same as the average energy loss of the jet as a whole, not more.
We saw in Section 5.2 that this results in a reduction in the total energy lost by
the jet, which gets compensated for by an increase in $\aSC$, the
hybrid model parameter that controls the magnitude of the rate of energy loss for all
partons in the jet, in order
to maintain the agreement between the prediction of the model for jet $R_{AA}$ and the
experimentally measured value of this quantity.
Consequently, when we turn on $\lres$ we end up with jets within which the few hardest hadrons at the
core of the jet lose modestly more energy and within which hadrons with moderate
energies loses modestly less energy, making them more likely to survive 
and populate the jet at larger angles relative to its
core.  We illustrated both effects by computing model predictions
for fragmentation functions and for the jet shape observable in Section 5.3, 
showing that both depend on $\lres$ in the expected fashion.
We defer the study of single hadron $R_{AA}$ and its potential sensitivity to $\lres$ to 
future work, but note here that turning on $\lres$ should push it somewhat downward 
since it is controlled by the hardest parton in each jet.

As we noted in Section 5.3, including the effects of resolution pushes the predictions
of the hybrid model for fragmentation functions and jet shapes toward their 
experimentally measured values, but the effects
are modest in magnitude and the discrepancies between the predictions of the model 
and the data that were highlighted in Ref.~\cite{Casalderrey-Solana:2016jvj} remain.
This provides indirect support for the alternative suggestion for the
origin of these discrepancies, namely that they are telling us that the
treatment of the backreaction of the plasma to the jets that
we have implemented, following Ref.~\cite{Casalderrey-Solana:2016jvj}, is 
not sufficient in detail.
In Ref.~\cite{Casalderrey-Solana:2016jvj}, this hypothesis was investigated
further via computing the hybrid model predictions for several missing-$p_T$ 
observables, comparing them to experimental data, and finding significant 
discrepancies, in particular for hadrons with 2~GeV~$<p_T<$~4~GeV.
In Section 5.4 we carried out this computation with $\lres$ turned on,
and found that the model predictions for these missing-$p_T$ observables
with $\lres$ set to $2/\pi T$ are very similar to those with $\lres=0$, with the differences
almost too small to see relative to the interesting remaining discrepancies between the model
predictions and data.
This provides direct support for the conclusion that these discrepancies are not
due to the effects of resolution and  indirect support for the
suggestion that they are telling us that the wakes left in the plasma by passing jets
do not have time to equilibrate fully, since if they had done so the wakes would have yielded
more hadrons with $p_T<2$~GeV and fewer hadrons with  2~GeV~$<p_T<4$~GeV reconstructed within the jets, as in the model calculation.
This result motivates a full event-by-event
hydrodynamic treatment of the wakes that goes beyond the linear approximations
that we have employed and suggests that a comparison between the predictions of such
a treatment and data could provide a snapshot of a non-equilibrium disturbance of
the strongly coupled plasma as it is equilibrating, but not
yet equilibrated.  These conclusions confirm the importance of 
further measurements of experimental observables
that are sensitive to the distribution of partons within the jet shower as a function of
energy and angle, which have increasing statistical precision, which are increasingly
differential, and which employ jets produced back-to-back with photons or $Z$-bosons
as well as in dijet pairs.

\acknowledgments

We thank J. Casalderrey-Solana, D.~Gulhan, G.~Milhano, and K.~Zapp for helpful discussions. We also thank J. Balewski, M. Goncharov, and C. Paus for their assistance with computing.
ZH and KR acknowledge the hospitality of the CERN Theory Group.
DP acknowledges the hospitality of the MIT Center for Theoretical Physics. 
The work of ZH was supported by the Undergraduate Research Opportunity Program at MIT.
The work of DP was supported by the U.S.~National Science Foundation within the framework of the JETSCAPE collaboration, under grant number ACI-1550300, and also
by grants FPA2013-46570 and MDM-2014-0369 
of ICCUB (Unidad de Excelencia `Mar\'ia de Maeztu') 
from the Spanish MINECO,  by grant 2014-SGR-104 from the 
Generalitat de Catalunya,
and by the Consolider CPAN project.
The work of KR was supported by the U.S.~Department of Energy under grant Contract
Number DE-SC0011090.

\end{document}